\documentclass[
pra,twocolumn,
showpacs,
floatfix]
{revtex4}

\usepackage{latexsym}
\usepackage{amssymb, amsmath}
\usepackage{exscale}
\usepackage{bm, graphics}
\usepackage{graphicx}

\newcommand{\D}{\mbox{\rm d}}
\newcommand{\uhb}[1]{\underline{\hat{\bf{#1}}}}
\newcommand{\hb}[1]{\hat{\bf{#1}}}
\newcommand{\uh}[1]{\underline{\hat{#1}}}

\begin{document}

\title{Cavity-assisted spontaneous emission
as a single-photon source: Pulse shape
and efficiency of one-photon Fock state preparation}

\author{M. Khanbekyan}
\email[E-mail address: ]{mkh@tpi.uni-jena.de}

\author{D.-G. Welsch}
\affiliation{Theoretisch-Physikalisches Institut,
Friedrich-Schiller-Universit\"at Jena, Max-Wien-Platz 1,
D-07743 Jena, Germany}

\author{C. Di Fidio}

\author{W. Vogel}
\affiliation{Arbeitsgruppe Quantenoptik, Institut f\"ur Physik,
Universit\"at Rostock, D-18051 Rostock, Germany}

\date{\today}

\begin{abstract}
Within the framework of exact
quantum electrodynamics in dispersing and absorbing media,
we have studied the quantum state of the radiation emitted from
an initially in the upper state prepared
two-level atom in a high-$Q$ cavity,
including the regime where the emitted photon 
belongs to a wave packet that simultaneously covers the
areas inside and outside the cavity.
For both continuing atom--field interaction
and short-term atom--field interaction, we
have determined the
spatio-temporal shape of the excited outgoing wave packet
and calculated the efficiency of the
wave packet
to carry a one-photon Fock state.
Furthermore, we have made contact with quantum noise
theories where the intracavity field and the field outside
the cavity are regarded as approximately representing
independent degrees of freedom such that two
separate Hilbert spaces can be introduced.

\end{abstract}


\maketitle

\section{Introduction}
\label{introduction}

The interaction of a single atom with a
quantized radiation-field mode in a high-$Q$ cavity
has played an important role
not only due to its conceptual relevance,
but also because it serves as a
basic ingredient in various
schemes in quantum optics and related fields
such as quantum information
science (for a review see, e.\,g.,
Refs.~\cite{walther:617, walther:1325}).
In fact, cavity quantum electrodynamics (QED)
has allowed the generation and processing of nonclassical
radiation and offered novel radiation sources such as the
single-atom maser \cite{meschede:551, brune:1899}
and laser \cite{ginzel:732, smith:2730,agarwal:1737, boozer:023814}
and the ion-trap laser
\cite{loeffler:3923,meyer:1099,difidio:013811,filho:013808}.
In this context, quantum control of single-photon
emission from an atom in a cavity
for generating single-photon Fock states on demand
has been an essential prerequisite \cite{monroe:238}.
In particular, single-photon Fock state generation
of high efficiency, as boosted by the well-pronounced line
spectra of cavity fields, has been
a key requirement in various applications such as
quantum cryptography \cite{bennett:2724, luetkenhaus:52304}
or quantum networking
for distribution and processing of quantum information
\cite{cirac:3221, knill:461}.
Recently, single-photon sources operating
on the basis of adiabatic passage with just one atom
trapped in a \mbox{high-$Q$} optical cavity has been realized
\cite{parkins:3095,mckeever:1992,hennrich:4872}.
In this way, generation of single-photons of
known circular polarization has been possible \cite{wilk:063601}.
Moreover, adjustment of the spatio-temporal profile of
single-photon pulses has been achieved
\cite{kuhn:067901, keller:1075}.

In view of the very wide-spread applications of
cavity-assisted single-photon sources, it is of great
importance to carefully study the quantum state of
the field escaping from a cavity.
Let us consider the simplest case of a two-level atom that
near-resonantly interacts with a narrow-band
cavity-field mode. On a time scale that is
sufficiently short compared to the inverse
bandwidth of the mode, the radiative and non-radiative
cavity losses may be disregarded, and the
atom--field dynamics can be described by the
familiar Jaynes-Cummings model~\cite{jaynes:89}.
Clearly, for longer times, the atom--cavity system cannot longer
be regarded as being a closed system, and the losses must
be taken into account. Since the wanted outgoing field
represents, from the point of view of the atom--cavity system,
radiative losses, the study of the input--output problem
necessarily requires inclusion in the theory of the effects
of losses.

There are primarily two approaches to the
problem, namely the approach based on
quantum noise theory and the one based
on macroscopic QED.
In quantum noise theory (QNT), the fields inside and
outside a cavity are regarded as representing independent
degrees of freedom, as it would be the case if the
cavity were bounded by perfectly reflecting walls
\cite{collett:1386, gardiner:3761, gardiner}.
Accordingly, QNT is based on discrete and continuous
mode expansions of the fields inside and outside the
cavity, respectively, so that inside- and outside-field
operators can be regarded as being commuting quantities.
In order to \emph{a posteriori} take into
account the input--output coupling due to non-perfectly
reflecting mirrors, each intracavity mode is linearly coupled
to the continuum of the external modes, which is regarded as
playing the role of a dissipative system. Its effect on the
intracavity modes is treated in Markovian approximation, leading to
quantum Langevin equations for the intracavity-mode operators,
where the incoming external field gives rise to the operator Langevin
forces therein.
Additionally to the radiative losses associated with a normally
wanted input--output coupling, there are always unwanted
losses such as absorption and scattering losses.
They can be straightforwardly included in the quantum
Langevin equations by coupling the intracavity modes to
additional dissipative systems and treating these interactions
in Markovian approximation
\cite{khanbekyan:043807}.
Alternatively to the concept of quantum Langevin equations,
dissipation can be described by using the concept of master
equations (see, e.\,g. Refs~\cite{haake,louisell,davies}).
There are different methods to solve master
equations, for example, the method of quantum trajectories
(see, e.\,g. Refs.~\cite{dalibard:580,dum:4382,carmichael}).
In order to find the field escaping from the cavity,
the quantum Langevin equations (or the equivalent master
equations) are completed with input--output relations,
which relate the output field to the input field
and the intracavity field.

In macroscopic QED, the system is described on the basis of the
respective macroscopic Maxwell equations \cite{knoell:1,vogel}.
By starting from an ordinary continuous-mode expansion of the
electromagnetic field in the presence of nonabsorbing
linear media, it can be shown that in some approximation
a description of the fields inside and outside a cavity in terms
of quantum Langevin equations and input--output relations,
respectively, as used in QNT can indeed be given
\cite{knoell:543, plank:1791}.
In another version of QED~\cite{viviescas:013805, viviescas:211},
solutions of Maxwell's equations are constructed 
by using Feshbach's projection formalism~\cite{feshbach:287}.
By means of the appropriately chosen
boundary conditions a decomposition of the field
can be performed which renders it possible
to catch up with the description of the
cavity system within the framework of QNT.
The method can be extended also to the case of
overlapping cavity modes in the case of lower $Q$~values.
An alternative approach is based on an
expansion of the fields inside and outside a cavity
into nonorthogonal Fox-Li modes \cite{dutra:063805}.
In this approach, however, the interaction energy
between the cavity modes and the external modes vanishes
and the input--output coupling arises from the non-zero
commutator between the fields inside and outside the cavity.

Allowing for dispersing and absorbing media,
one can also use macroscopic QED to
include in the theory the effect of
unwanted losses \cite{khanbekyan:053813}, which,
in agreement with QNT, can be shown to become manifest in
additional damping terms and the associated fluctuation forces
in the quantum Langevin equations.
In contrast, inclusion in the input--output relations
of the effect of unwanted losses is not straightforward
since it cannot be deduced
from the interaction Hamiltonians used in QNT
\cite{khanbekyan:053813,semenov:033803,semenov:013807}.
Particularly, input--output relations suggested by QNT
do not describe the effect of unwanted losses
on the output field which is induced by the reflected
input field.

The input--output relations can be used to introduce a
many-mode characteristic function
of the quantum state of the output field,
from which the quantum state can be inferred
in terms of phase-space functions.
In Ref.~\cite{khanbekyan:053813}, explicit results are given
for the case, when the time necessary to prepare
an intracavity mode in some quantum
state is sufficiently short compared to the decay time
of this mode so that the preparation process may
be disregarded and instead, an initial condition can be set
for the quantum state of the intracavity mode \cite{khanbekyan:053813}.
Furthermore, it is assumed that the fields
inside and outside the cavity can be regarded as
being effectively commuting quantities at equal times.
In this way, the mode structure of the output field
is determined, and the Wigner
function of the quantum state of the relevant
output mode---the one that is related to the
excited intracavity-mode---is expressed in terms of the
Wigner functions of the quantum states of the
intracavity mode, the incoming field, and the
radiationless dissipative system.
Needless to say that the simplifying assumption
of short preparation limits the scope of the results in general.

In the present paper we generalize the approach based on
macroscopic QED in dispersing and absorbing media,
with the aim to renounce the approximation
that the electromagnetic fields inside and outside
a cavity represent independent degrees of freedom. Instead, 
we treat the electromagnetic field as an entity. Further, 
we include in the theory the preparation process
in order to go beyond the regime of short-time preparation.
We work out the theory for the case where
the quantum state of the outgoing field results from the
resonant interaction of the electromagnetic field with
a single two-level atom initially prepared in the upper 
state. Considering a source-quantity representation
of the electromagnetic field and treating the 
atom--field interaction in rotating-wave approximation,
we examine the mode structure of the outgoing field
as well as the efficiency of the excited outgoing mode 
to carry a single-photon Fock state.
We finally compare the results with the ones
obtained within the framework of QNT.

The paper is organized as follows. The basic 
equations for the resonant interaction of a two-level
atom with a cavity-assisted electromagnetic field
are given in Sec.~\ref{sec3.1}. In Sec.~\ref{sec5}, 
the Wigner function of the quantum state of the excited
outgoing wave packet and the shape of the wave packet
are studied for different atom--field interaction
times and a comparison with QNT is made.
A summary and some concluding remarks are given
in Sec.~\ref{summary}.


\section{Basic equations}
\label{sec3.1}

The starting point of QED are the macroscopic 
Maxwell equations for the medium-assisted electromagnetic 
field coupled to the equations of motion of the
active atomic sources considered, where the
effect of the medium is described by appropriately chosen
constitutive equations. In particular, the effect of a
locally responding, inhomogeneous, linear dielectric, which we
will focus on throughout the paper, can be described by a spatially
varying (relative) permittivity $\varepsilon(\mathbf{r},\omega)$,
which is a complex function of frequency,
\begin{equation}
    \label{1.2}
      \varepsilon({\bf r},\omega) = \varepsilon'({\bf r},\omega)
      + i\varepsilon''({\bf r},\omega),
\end{equation}
with the real and imaginary parts $\varepsilon'({\bf r},\omega)$
and $\varepsilon''({\bf r},\omega)$, respectively, being related
to each other via the Kramers--Kronig relations.


\subsection{Quantization scheme}
\label{sec3.2}

To be more specific, let us consider $N$ atoms
that interact with the electromagnetic field
in the presence of a dielectric medium.
Applying the multipolar-coupling scheme in electric dipole
approximation, we may write the Hamiltonian
that governs the temporal evolution
of the overall system, which consists of the electromagnetic
field, the dielectric medium (including the dissipative degrees
of freedom), and the atoms coupled to the field,
in the form \cite{knoell:1, vogel}
\begin{align}
   \label{1.1}
        \hat{H} =
&
        \int\! \D^3{r} \int_0^\infty\! \D\omega
      \,\hbar\omega\,\hb {f}^{\dagger}({\bf r },\omega)\cdot
      \hb{ f}({\bf r},\omega)
\nonumber\\&
+
         \sum _A \sum _k
        \hbar \omega _{Ak} \hat{S} _{Akk}
        -\sum _A
        \hb{ d}_A\cdot
        \hb{E}({\bf r}_A).
\end{align}
In this equation, the first term is the Hamiltonian of
the \mbox{field--me}\-dium system, where the bosonic
fields \mbox{$\hb{ f}({\bf r},\omega)$}
and \mbox{$\hb{f}^\dagger({\bf r},\omega)$},
\begin{align}
    \label{1.3}
&      \bigl[\hat{f}_{\mu} ({\bf r}, \omega),
      \hat{f}_{\mu'} ^{\dagger } ({\bf r }',  \omega ') \bigr]
      = \delta _{\mu \mu'}\delta (\omega - \omega  ')
      \delta ^{(3)}({\bf r} - {\bf r }') ,
\\
\label{1.3-1}
&\bigl[\hat{f}_{\mu} ({\bf r}, \omega),
      \hat{f}_{\mu'} ({\bf r }',  \omega ') \bigr]
= 0,
\end{align}
play the role of the canonically conjugate system variables.
The second term is the Hamiltonian of the atoms, where
the $\hat{S}_{Ak'k}$ are the atomic flip operators for
the $A$th atom,
\begin{equation}
   \label{1.5}
   \hat{S} _{Ak'k} =
   | k'\rangle_{\!A} {_A}\!\langle k |
,
\end{equation}
with the $|k\rangle_{\!A}$ being the energy eigenstates
of the $A$th atom. Finally, the last term is the atom--field 
coupling energy, where
\begin{equation}
   \label{1.7}
    \hb{ d}_A = \sum _{kk'}
    {\bf d} _{Akk'}  \hat{S} _{Akk'}
\end{equation}
is the electric dipole moment of the $A$th atom
($ {\bf d} _{Akk'}$ $\!=$ $\!{_A}\!\langle k|
\hb{ d}_{\!A} | k' \rangle_A$), and the 
medium-assisted electric field $\hb{E}({\bf r})$ 
can be expressed in terms of the variables
$\hat{\mathbf{f}}(\mathbf{r},\omega)$ and
$\hat{\mathbf{f}}^\dagger(\mathbf{r},\omega)$ as
follows~\footnote{Note that in the
   multipolar-coupling scheme,
   $\hat{\mathbf{E}}(\mathbf{r})$ may behave like a
   displacement field
   with respect to the atomic
   polarization field.}:
\begin{equation}
\label{1.9}
\hb{E}({\bf r}) = \hb{E}^{(+)}({\bf r})
        +\hb{E}^{(-)}({\bf r}),
\end{equation}
\begin{equation}
\label{1.10}
\hb{E}^{(+)}({\bf r}) = \int_0^\infty \D\omega\,
      \uhb{E}({\bf r},\omega),
\quad
\hb{E}^{(-)}({\bf r}) =
[\hb{E}^{(+)}({\bf r})]^\dagger,
\end{equation}
\begin{multline}
      \label{1.11}
      \uhb{ E}({\bf r},\omega)
\\[.5ex]
=
i \sqrt{\frac {\hbar}{\varepsilon_0\pi}}\,
\frac{ \omega^2}{c^2}
      \int \D^3r'\sqrt{\varepsilon''({\bf r}',\omega)}\,
      \mathsf{G}({\bf r},{\bf r}',\omega)
      \cdot\hb{f}({\bf r}',\omega),
\end{multline}
where the classical (retarded)
Green tensor $\mathsf{G}({\bf r},{\bf r}',\omega)$
is the solution to the equation
\begin{equation}
      \label{1.13}
      \bm{\nabla}\!\times\!
    \bm{\nabla}\!  \times \mathsf{G}  ({\bf r }, {\bf r }', \omega)
      - \frac {\omega ^2 } {c^2} \,\varepsilon ( {\bf r } ,\omega)
      \mathsf{G}  ({\bf r}, {\bf r }', \omega)
      =  \bm{\delta} ^{(3)}  ({\bf r }-{\bf r }')
      \end{equation}
and satisfies the boundary condition at infinity, i.\,e.,
$\mathsf{G}({\bf r},{\bf r}',\omega)\to 0$ if
$|\mathbf{r}-\mathbf{r}'|\to\infty$.


\subsection{Two-level atom in a cavity}
\label{sec3.3}

\begin{figure}[t]
\includegraphics[width=.9\linewidth]{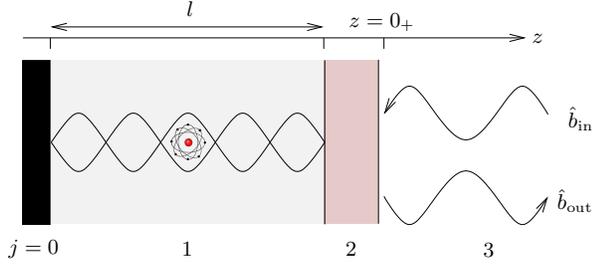}
\caption{\label{fig}Scheme of the system. The fractionally 
transparent mirror of the cavity (region~2) is modeled by
a dielectric plate, and the atom inside the cavity (region~1)
can be embedded in some dielectric medium.
}
\end{figure}

Let us focus on a single two-level atom at position $z_A$
inside a high-$Q$ cavity and use the model of Ley and Loudon
\cite{ley:227}, i.\,e., a one-dimensional cavity in $z$-direction
which is bounded by a perfectly reflecting mirror at 
the left-hand side and a
fractionally transparent mirror at the 
right-hand side, and a
linearly polarized electromagnetic field
which propagates along the $z$~axis (Fig.~\ref{fig}).
Regarding the cavity as a multi-layer dielectric system
and restricting our attention to resonant
atom--field interaction, we may start from the
one-dimensional version of the Hamiltonian (\ref{1.1})
and apply the rotating-wave approximation to the
atom--field interaction, leading to
\begin{multline}
   \label{1.15}
        \hat{H} =
        \int\! \D z\int_0^\infty\! \D\omega
      \,\hbar\omega\,\hat {f}^{\dagger}(z, \omega)
      \hat{f}(z, \omega)
+ \hbar \omega _{0} \hat {S}_{22}
\\
  -\left[
        d_{21}\hat{S}_{12}^{\dagger}
            \hat{E}^{(+)}(z_A)
            +
\mbox{H.c.} \right]
\end{multline}
($\hat {S}_{k'k}$ $\equiv$ $\!\hat {S}_{Ak'k}$),
where $\omega_0$ is the atomic transition frequency.

In what follows we assume that
the atom is initially (at time $t$ $\!=$ $\!0$)
prepared in the upper state $|2\rangle$ and the rest of
the system, i.\,e., the combined system that consists of the
electromagnetic field and the cavity, is in the ground
state $|\{0\}\rangle$. We may therefore expand the state
vector of the overall system at a later time $t$ ($t\ge 0$) as
\begin{multline}
\label{1.17}
  |\psi(t)\rangle =
  C_2(t)e^{-i\omega_{0} t}
  |2\rangle|\!\left\lbrace0\right\rbrace\!\rangle
\\[.5ex]
   +\int \D z\, \int_0^\infty \D \omega\,
   C_1(z, \omega, t) e^{-i\omega t}
   |1\rangle\hat{f}^{\dagger}(z, \omega)
   |\!\left\lbrace0\right\rbrace\!\rangle ,
\end{multline}
where $|1\rangle$ is the lower atomic state,
and $\hat{f}^{\dagger}(z, \omega)|\{0\}\rangle$
is a single-quantum excited state of the combined
field--cavity system. It is not difficult to prove 
that the Schr\"odinger equation for $|\psi(t)\rangle$ 
then leads to the following system of differential 
equations for the probability amplitudes
$C_2(t)$ and $C_1(z, \omega, t)$:
\begin{multline}
  \label{1.19}
 \dot {C_2} =
      -\frac{d_{21}}{\sqrt{\pi \hbar \epsilon _0  \mathcal{A}}}
      \int_0^\infty\! \D\omega\, \frac{\omega ^2}{c^2}
      \int \D z
      \sqrt{\varepsilon''(z,\omega)}\,
\\[.5ex]
\times
      G(z_A, z,\omega)
      C_1(z, \omega, t)
      e^{-i (\omega - \omega_0)t}
     ,
\end{multline}
\begin{multline}
  \label{1.21}
 \dot {C_1}(z, \omega, t) =
      \frac{d_{21}^*}{\sqrt{\pi \hbar \epsilon _0  \mathcal{A}}}
      \frac{\omega ^2}{c^2}
       \sqrt{\varepsilon''(z,\omega)}\,
\\[.5ex]
\times
      G^*(z_A, z,\omega)
      C_2(t)
      e^{i (\omega - \omega_0)t}
\end{multline}
($\mathcal{A}$, mirror area).
Substituting the formal solution of Eq.~(\ref{1.21})
[with the initial condition \mbox{$C_1(z, \omega, 0)$ 
$\!=$ $\!0$}],
\begin{multline}
  \label{1.22}
 C_1(z, \omega, t) =
      \frac{d_{21}^*}{\sqrt{\pi \hbar \epsilon _0  \mathcal{A}}}
      \frac{\omega ^2}{c^2}
       \sqrt{\varepsilon''(z,\omega)}\,
            G^*(z_A, z,\omega)
\\[.5ex]
\times
\int_0^t \D t'\,
      C_2(t')
      e^{i (\omega - \omega_0)t'}
,
\end{multline}
into Eq.~(\ref{1.19}) and employing the integral relation
\begin{equation}
  \label{1.23}
  \mathrm{Im}\,G
  (z_1, z_2,\omega)
  = \frac {\omega ^2} {c^2}
  \!\int\!  \D z\,
  \varepsilon'' (z, \omega)
  G(z_1, z, \omega)G^*(z_2, z, \omega),
\end{equation}
we obtain the integro-differential equation
\begin{equation}
  \label{1.25}
    \dot {C}_2(t)
    = \int_0^t \! \D t'\,
    K(t-t')
    C_2(t'),
\end{equation}
where the integral kernel $K(t)$ reads
\begin{equation}
  \label{1.27}
K(t) =
        -\frac{|d_{21}|^2}
              {\pi \hbar \epsilon _0  \mathcal{A}}
      \int_0^\infty\! \D\omega\, \frac{\omega ^2}{c^2}\,
       e^{-i (\omega - \omega_0)t}
       \mathrm{Im}\,G(z_A, z_A,\omega).
\end{equation}

We now recall that the spectral response of the cavity
field is determined by the Green function $G(z,z',\omega)$
(see Appendix~\ref{app:1}). For a sufficiently high-$Q$ cavity,
the excitation spectrum effectively turns into
a quasi-discrete set of lines of mid-frequencies $\omega_k$
and widths $\Gamma_k$, according to the poles of the
Green function at the complex frequencies
\begin{equation}
  \label{1.30}
    \Omega _{k} = \omega_{k}
      - {\textstyle\frac {1} {2}}i\Gamma _{k},
\end{equation}
where the line widths are much smaller than the
line separations,
\begin{equation}
\label{1.30-1}
\Gamma_k \ll {\textstyle\frac{1}{2}}(\omega_{k+1} - \omega_{k-1}).
\end{equation}
In this case, we can divide the $\omega$~axis into intervals
\mbox{$\Delta_k$ $\!=$ $[\frac{1}{2}(\omega_{k-1}$ $\!+$
$\!\omega_k),\frac{1}{2}(\omega_k$ $\!+$ $\!\omega_{k+1})]$}
and rewrite Eq.~(\ref{1.27}) as
\begin{multline}
  \label{1.32}
K(t) =
\\[.5ex]
        -\frac{|d_{21}|^2}
              {\pi \hbar \epsilon _0  \mathcal{A}}
     \sum_k
     \int _{\Delta_k}\!
     \D\omega\, \frac{\omega ^2}{c^2}\,
       e^{-i (\omega - \omega_0)t}
       \mathrm{Im}\,G(z_A, z_A,\omega).
\end{multline}
Inserting Eq.~(\ref{1.32}) into Eq.~(\ref{1.25}), we obtain
\begin{multline}
  \label{1.26}
    \dot {C}_2(t)
    =
     -\frac{|d_{21}|^2}
     {\pi \hbar \epsilon _0  \mathcal{A}}
     \sum_k
    \int_0^t \! \D t'\,
      \int _{\Delta_k}\!
     \D\omega\, \frac{\omega ^2}{c^2}
\\[.5ex]
\times
      e^{-i (\omega - \omega_0)(t-t')}
       \mathrm{Im}\,G(z_A, z_A,\omega) C_2(t').
\end{multline}

To take into account the cavity-induced shift $\delta\omega$
of the atomic transition frequency, we make the ansatz
\begin{align}
  \label{1.34}
C_2(t) = e^{i\delta\omega t}\tilde{C}_2 (t)
\end{align}
and find from Eq.~(\ref{1.26})
\begin{multline}
  \label{1.45}
    \dot  {\tilde C}_2(t)
    =
    -i\delta\omega
 {\tilde C}_2 (t)
     -\frac{|d_{21}|^2}
     {\pi \hbar \epsilon _0  \mathcal{A}}
     \sum_k
    \int_0^t \! \D t'
      \int _{\Delta_k}\!
     \D\omega\, \frac{\omega ^2}{c^2}
\\[.5ex]
\times
      e^{-i (\omega - {\tilde \omega}_0)(t-t')}
       \mathrm{Im}\,G(z_A, z_A,\omega)  {\tilde C}_2 (t'),
\end{multline}
where
\begin{equation}
  \label{1.42}
   {\tilde \omega}_0 = \omega_0 - \delta\omega
.
\end{equation}
Let us assume that the atomic transition frequency $\omega _0$
is nearby a cavity resonance frequency, say $\omega_k $,
so that strong atom--field coupling may be realized.
Then, the exponential \mbox{$\exp[-i(\omega$ $\!-$
$\!{\tilde \omega} _0)(t$ $\!-$ $\!t')]$}
can be regarded, with respect to time,
as being rapidly oscillating in all
the off-resonant terms with $k'\neq k$ in the sum
in Eq.~(\ref{1.45}), and the time integrals in these 
terms can be performed in Markov approximation.
That is, replacing ${\tilde C}_2(t')$
in the off-resonant terms with ${\tilde C}_2(t)$
and confining ourselves to the times large compared to
$\Delta_k^{-1}$, we may identify $\delta\omega$ with
\begin{equation}
  \label{1.36}
    \delta\omega
= -\frac{|d_{21}|^2}
              {\pi \hbar \epsilon _0  \mathcal{A}}
              \sum_{k'\neq k}
               \int _{\Delta_{k'}}\! \D\omega\,
               \frac{\omega ^2}{c^2}\,
               \frac{ \mathrm{Im}\,G (z_A, z_A,\omega)}
               {{\tilde \omega}_0 - \omega}.
\end{equation}
Then, from Eq.~(\ref{1.45}) we can see that ${\tilde C}_2 (t)$
obeys the integro-differential equation
\begin{equation}
  \label{1.38}
    \dot {{\tilde C}}_2(t)
    = \int_0^t \! \D t'\,
    {\tilde K}(t-t')
    {\tilde C}_2(t'),
\end{equation}
where the kernel function $\tilde{K}(t)$ reads
\begin{equation}
  \label{1.40}
{\tilde K}(t) =
        -\frac{|d_{21}|^2}
              {\pi \hbar \epsilon _0  \mathcal{A}}
      \int_{\Delta_k}\! \D\omega\, \frac{\omega ^2}{c^2}\,
       e^{-i (\omega - {\tilde \omega}_0)t}
       \mathrm{Im}\,G(z_A, z_A,\omega)
.
\end{equation}

In fact, Eq.~(\ref{1.36}) can only be used to
calculate the \mbox{$\mathbf{r}_A$-de}\-pen\-dent part
of the frequency shift, i.\,e., the cavity-induced part
which arises from the scattering part of the Green
function [for the decomposition of the Green function
in bulk and scattering parts, see Eq.~(\ref{app2.1})].
The $\mathbf{r}_A$-independent part which arises from the bulk part
of the Green function and which is not cavity-specific would diverge,
particularly because of the dipole approximation made.
Since this part can be thought of as being already included in
the definition of the transition frequency $\omega_0$, we can
focus on the $\mathbf{r}_A$-dependent part.
Inserting the scattering part of the Green function
into  Eq.~(\ref{1.36}), we derive (Appendix \ref{app:2})
\begin{multline}
  \label{1.43}
    \delta\omega
    = - \sum_{k'
    }\frac{
    \alpha _{k'}
  }
  {4|{\tilde \omega}_0 - \Omega _{k'}|^2}
\\[.5ex]
\times
\left[
    {\tilde \omega}_0 \omega _{k'}-|\Omega _{k'}|^2
      -
     \frac{\tilde{\omega}_0\Gamma_{k'}}{4\pi}\,
      \ln\!\left(\frac{\omega _{k'}}{\omega _0}\right)
\right]
\end{multline}
with
\begin{equation}
  \label{1.39}
   \alpha_k  =  \frac{4|d_{21}|^2 }
    {\hbar \epsilon _0  \mathcal{A}|n_1
(\Omega_k)
    |^2 l}\,
    \sin^2
    [\omega_k |n_1
(\Omega_k)
    |
    z_A/c]
,
\end{equation}
where $l$ is the length of the cavity
(Fig.~\ref{fig}) and $n_1(\omega)$
is the (complex) refractive
index of the medium inside the cavity.

To calculate the kernel function $\tilde{K}(t)$, Eq.~(\ref{1.40}),
we note that, within the approximation scheme used, the frequency
integration can be extended to $\pm\infty$. Employing the Green 
function as given by Eq.~(\ref{app.1}) and approximating 
$\tilde{K}(t)$ by its leading-order contribution, we derive
\begin{equation}
  \label{1.29}
 {\tilde K}(t) =
      -
      {\textstyle\frac {1} {4}}
      \alpha_k
      \Omega_k
        e^{-i (\Omega_k - {\tilde \omega}_0)t}
        .
\end{equation}
Having solved Eq.~(\ref{1.38}) and calculated $\tilde{C}_2(t)$,
we may eventually calculate $C_1(t)$ according to Eq.~(\ref{1.22}):
\begin{multline}
  \label{1.44}
     C_1(z, \omega, t) =
      \frac{d_{21}^*}{\sqrt{\pi \hbar \epsilon _0  \mathcal{A}}}
      \frac{\omega ^2}{c^2}\,
      G^*(z_A, z,\omega)
\\[.5ex]
\times
       \int_0^t  \D t'\,
       \sqrt{\varepsilon''(z,\omega)}\,
      {\tilde C}_2(t')
      e^{i (\omega - {\tilde \omega}_0)t'}.
\end{multline}


\section{Quantum state of the outgoing field}
\label{sec5}

For the sake of transparency, let us restrict our attention
to the case where the cavity is embedded in free space.
Inserting the Green tensor as given by Eq.~(\ref{app.1})
in the one-dimensional version of Eq.~(\ref{1.11})
and decomposing the electric field outside the cavity into
incoming and outgoing fields, we may represent the
outgoing field, for example, at the point 
$z$ $\!=$ $\!0^+$ (cf.~Fig.~\ref{fig}) as
\cite{khanbekyan:053813}
 \begin{multline}
      \label{5.1}
     \uh{ E}_{\mathrm{out}}(z,\omega)
       \bigr|_{z=0^+}
      = i \sqrt{\frac {\hbar }{\epsilon _0\pi \mathcal{A}}}\,
      \frac{\omega^2}{c^2}
\\[.5ex]
\times
      \int \D z'\sqrt{\varepsilon''(z',\omega)}\,
      G_{\rm out}(0^+,z',\omega)
      \hat{f}(z',\omega),
\end{multline}
where, according to Eq.~(\ref{1.3}), the commutation relation
\begin{equation}
    \label{5.2}
     \left[\hat{f}(z, \omega),
      \hat{f}^{\dagger } (z',  \omega ') \right]
      = \delta (\omega - \omega  ')
      \delta (z - z')
\end{equation}
holds. For the following it will be useful to introduce
the bosonic operators
\begin{equation}
      \label{5.3}
      \hat{ b}_{\mathrm{out}}  (\omega)
      = 2\,\sqrt{\frac{\varepsilon_0 c \pi\mathcal{A}}{\hbar\omega}}
     \,\left.\uh{ E}_{\mathrm{out}}(z, \omega)
     \right|_{z=0^+}
.
\end{equation}
It is not difficult to prove that
\begin{equation}
  \label{5.5}
  \left[\hat{b}_{\mathrm{out}}(\omega),
       \hat{b}_{\mathrm{out}}^\dagger(\omega')\right]
       =
\delta (\omega - \omega ').
\end{equation}


\subsection{Wigner function}
\label{sec5.1}

To calculate the quantum state of the outgoing field,
we start from the multimode characteristic functional
\cite{khanbekyan:043807}
\begin{multline}
\label{5.7}
C_{{\rm out}}[\beta(\omega),t]
\\[.5ex]
         =
         \left
          \langle
           \psi(t)
         \right|
         \exp\!\left[
         \int_0^{\infty}\!
         \D\omega\,
         \beta(\omega) \hat{b}^{\dagger}_{\mathrm {out}}(\omega)
         - \mathrm{H.c.}\right]\!
       \left|
       \psi(t)
        \right
        \rangle
,
\end{multline}
i.\,e., the characteristic functional of the Wigner
functional. Applying the Baker--Campbell--Hausdorff formula
and recalling the commutation relation (\ref{5.5}),
we may rewrite $C_{{\rm out}}[\beta(\omega),t]$ as
\begin{multline}
\label{5.8}
C_{{\rm out}}[\beta(\omega),t]=
\exp\left[ -{\textstyle\frac{1}{2}}\int_0^{\infty}\!
  \D\omega\,
      |\beta (\omega)|^2
    \right]
\\[.5ex]\hspace{-4ex}\times
          \langle\psi(t)|
         \exp\!\left[
         \int_0^{\infty}\!
        \D\omega\,
         \beta(\omega) \hat{b}^{\dagger}_{\mathrm {out}}(\omega)
      \right]
\\[.5ex]\times
 \exp\!\left[
   -\!
         \int_0^{\infty}\!
        \D\omega\,
         \beta^*(\omega) \hat{b}_{\mathrm {out}}(\omega)
      \right]
        \!
       |\psi(t)\rangle
        .
\end{multline}
Note that Eq.~(\ref{5.8}) is quite generally valid as yet.

To evaluate $C_{{\rm out}}[\beta(\omega),t]$ for the
state $|\psi(t)\rangle$ as given by Eq.~(\ref{1.17}), we
first note that from Eq.~(\ref{1.17}) together with the 
commutation relation (\ref{5.2}) and the relation 
$\hat{f}(z, \omega)|\lbrace 0\rbrace\rangle$
$\!=$ $\!0$ it follows that
\begin{equation}
  \label{5.10}
   \hat{f}(z, \omega) |\psi(t)\rangle =
    C_1(z, \omega, t) e^{-i\omega t}
   |1\rangle\
   |\lbrace 0\rbrace\rangle.
\end{equation}
Hence, on recalling Eqs.~(\ref{5.1}) and (\ref{5.3}),
it can be seen that
\begin{equation}
  \label{5.6}
  \hat{ b}_{\mathrm{out}}  (\omega)
   |\psi(t)\rangle =
    F^*(\omega, t)
   |1\rangle\
   |\lbrace 0\rbrace\rangle,
\end{equation}
where
\begin{multline}
  \label{5.11}
    F(\omega, t)=
    -2i
   \sqrt{\frac{c}{\omega}}
    \frac{\omega^2}{c^2}
\\[.5ex]
\times
      \int \D z\,
       \sqrt{\varepsilon''(z, \omega)}\,
     G^*_{\rm out}(0^+,z,\omega)
     C_1^*(z,\omega, t)
     e^{i\omega t},
\end{multline}
with $C_1(z,\omega, t)$ being determined by Eq.~(\ref{1.44}).
With the help of Eq.~(\ref{5.3}) [together with
Eq.~(\ref{5.1})] and Eq.~(\ref{5.10}),
it is now not difficult to combine Eqs.~(\ref{5.8}) and
(\ref{5.6}) to obtain $C_{\mathrm{out}}[\beta(\omega),t]$ as
\begin{multline}
  \label{5.9}
   C_{\mathrm{out}}[\beta(\omega),t]=
    \exp\left[ -{\textstyle\frac{1}{2}}
\int_0^{\infty}\!
\D\omega\,
      |\beta (\omega)|^2
    \right]
\\[.5ex]
    \times
    \left[
      1-\left|\int_0^{\infty}\!
\D\omega\,
\beta(\omega) F(\omega, t)\right|^2
\right].
\end{multline}

To represent $C_{\mathrm{out}}[\beta(\omega),t]$ in a
more transparent form, we introduce a time-dependent
unitary transformation according to
\begin{align}
  \label{5.20}
&
        \beta(\omega)
         =
      \sum_i F_i^\ast
(\omega,t)
\beta_i(t),
\\[.5ex] \label{5.22}
&
\beta_i(t)
   =
   \int_0^{\infty}\!
    \D\omega\,
    F_i(\omega,t)
          \beta (\omega ) .
\end{align}
Inserting Eq.~(\ref{5.20}) in Eq.~(\ref{5.8}),
we may rewrite Eq.~(\ref{5.8}) as
[$C_{\mathrm{out}}[\beta(\omega),t]
\mapsto C_{\mathrm{out}}[\beta _i(t),t]$]
\begin{multline}
\label{5.16}
C_{{\rm out}}[\beta_i(t),t]
= \exp\!\left[-{\textstyle\frac{1}{2}}
\sum_i |\beta_i(t)|^2\right]
\\[.5ex]
\times
         \left\langle\psi(t)\right|
         \exp\!\left[\!
         \sum_i
         \beta_i (t)
\hat{b}_{\mathrm{out}\,i}^\dagger(t)
\right]
\\[.5ex]
\times
         \exp\!\left[-\!
         \sum_i
         \beta_i^\ast(t)
\hat{b}_{\mathrm{out}\,i}(t)
\right]
\left|\psi(t)\right\rangle
\!,
\end{multline}
where
\begin{equation}
  \label{5.15}
  \hat{b}_{\mathrm{out}\,i}(t)
      = \int_0^{\infty}\!
      \D\omega\,
    F_i(\omega,t)\hat{b}_{\mathrm {out}} (\omega)
\end{equation}
are the operators associated with nonmonochromatic modes 
$F_i(\omega,t)$ of the outgoing field, which are not 
yet specified. Note that
\begin{equation}
  \label{5.13}
\hat{b}_{\mathrm {out}} (\omega)
=
  \sum_i
  F_i^\ast(\omega,t)
  \hat{b}_{\mathrm{out}\,i}(t)
.
\end{equation}
Accordingly, we may rewrite Eq.~(\ref{5.9}) as
\begin{multline}
  \label{5.15-1}
   C_{\mathrm{out}}[\beta_i(t),t]=
    \exp\left[ -{\textstyle\frac{1}{2}}
    \sum_i
      |\beta_i(t)|^2
    \right]
\\[.5ex]
    \times
    \left[
      1-\left|\sum_i \beta_i(t)
     \int_0^{\infty}\!
     \D\omega\,
     F_i(\omega,t)F(\omega, t)\right|^2
    \right].
\end{multline}

We now choose
\begin{equation}
  \label{5.17}
    F_1(\omega,t)
    = \frac{F(\omega , t)}{\sqrt{\eta(t)}}\,,
\end{equation}
where, within the approximation scheme used,
\begin{equation}
 \label{5.19}
         \eta(t)
=
\int_0^{\infty}\!
         \D\omega\, |F(\omega ,t)| ^2
\simeq
\int_{-\infty}^\infty
         \D\omega\, |F(\omega ,t)| ^2,
 \end{equation}
with $F(\omega ,t)$ being given by Eq.~(\ref{5.11}).
In this way, from Eq.~(\ref{5.15-1}) we obtain
$C_{\mathrm{out}}[\beta_i(t),t]$ in a 'diagonal' form 
with respect to the nonmonochromatic modes:
\begin{equation}
  \label{5.21}
C_\mathrm{out}[\beta_i(t),t]
= C_1[\beta_1(t),t]\prod_{i\neq1} C_i[\beta_i(t),t],
\end{equation}
where
\begin{equation}
  \label{5.21-2}
 C_1(\beta,t) = e^{-|\beta|^2/2}
\left[1-\eta(t)|\beta|^2\right]
\end{equation}
and
\begin{equation}
  \label{5.21-3}
 C_i(\beta,t) = e^{-|\beta|^2/2} \quad (i\neq 1).
\end{equation}
Hence, the quantum state of the outgoing field
factorizes with respect to the nonmonochromatic 
modes $F_i(\omega,t)$.

The Fourier transform of $C_{\mathrm{out}}[\beta _i(t), t]$
with respect to the $\beta_i(t)$ then yields the (multi-mode)
Wigner function $W_{\mathrm{out}}(\alpha_i,t)$ sought,
\begin{multline}
\label{5.21-1}
W_{\mathrm{out}} (\alpha_i, t)
\\[.5ex]
= \frac{2}{\pi}
      \exp\!\left[-2\sum _i|\alpha _i|^2
    \right]
      \left[1-2\eta(t)(1-2 |\alpha_1|^2)\right]
,
\end{multline}
which can be rewritten as
\begin{equation}
\label{5.23}
W_{\mathrm{out}} (\alpha_i, t)
= W_1(\alpha_1,t)
\prod_{i\neq 1}
     W_i^{(0)}(\alpha _i, t),
\end{equation}
where
\begin{equation}
\label{5.23-1}
W_1(\alpha,t)
= [1-\eta(t)]W_1^{(0)}(\alpha)
     +\eta(t)W_1^{(1)}(\alpha),
\end{equation}
with $W_i^{(0)}(\alpha)$ and $W_i^{(1)}(\alpha)$,
respectively, being the Wigner functions of the vacuum
state and the one-photon Fock state of the
$i$th nonmonochromatic mode. As we see, the
mode labeled by the subscript \mbox{ $i$ $\!=$
$\!1$}---the excited outgoing mode---is
in the mixed state described by the Wigner function
$W_1(\alpha,t)$, which reveals that
$\eta(t)$ can be regarded as being the efficiency
to prepare the excited outgoing mode in
a one-photon Fock state.


\subsection{Continuing atom--field interaction}
\label{sec5.2}

The formulas derived above refer to the case of
continuing atom--field interaction. In particular, the
efficiency $\eta(t)$ of the excited outgoing mode 
being prepared in a one-photon Fock state,
as given by Eq.~(\ref{5.19}) together with Eq.~(\ref{5.11}),
refers to this case.
Its determination requires the calculation
of the probability amplitude $C_1(z,\omega,t)$,
which can be obtained from the probability amplitude
$\tilde{C}_2(t)$ according to Eq.~(\ref{1.44}).
In order to determine ${\tilde C}_2(t)$, we first
substitute Eq.~(\ref{1.29}) into Eq.~(\ref{1.38}) and
differentiate both sides of the resulting equation
with respect to time. In this way, we derive the following
second-order differential equation for $\tilde{C}_2(t)$:
\begin{equation}
  \label{1.31}
   \ddot{\tilde{C}}_2+i(\Omega_k
   -\tilde{\omega}_0)\dot{\tilde{C}}_2
   +
   {\textstyle\frac {1} {4}}
   \alpha_k \Omega _k
   \tilde{C}_2(t)=0 ,
\end{equation}
where
\begin{equation}
  \label{1.33}
  \zeta_k \equiv \rho_k -
  {\textstyle\frac {1} {2}}i
 \gamma_k
   =
   \sqrt{
        (
\Omega_k - {\tilde \omega}_0) ^2 +
   \alpha_k\Omega_k}\,
\end{equation}
[with $\Omega_k$ and $\alpha_k$ from Eqs.~(\ref{1.30})
and (\ref{1.39}), respectively].
The solution to Eq.~(\ref{1.31}) reads
 \begin{multline}
   \label{1.41}
      {\tilde C}_2(t) = e^{-i(\Omega _k - {\tilde \omega}_0)t/2}
\\\times
    \left[
    \cos(\zeta_k t/2)
    +i\frac{
\Omega_k - {\tilde \omega} _0
    } {\zeta_k}
    \sin(\zeta_k t/2)
   \right]
    .
\end{multline}
Note that when \mbox{$\rho_k $ $\!\gg $
$\!\frac{1}{2}(\Gamma_k$ $\!+$ $\!\gamma_k)$},
then damped vacuum Rabi oscillations of the 
upper-state occupation probability
$|\tilde{C}_2(t)|^2$ $\!=$ $\!|C_2(t)|^2$ 
[recall Eq.~(\ref{1.34})] are observed (Fig.~\ref{fig2}), 
where the vacuum Rabi frequency is given by
$R_k$ $\!=$ $\!\sqrt{\alpha_k\omega_k}$.


\subsubsection{One-photon-Fock-state extraction efficiency}
\label{sec5.2.1}

To calculate $\eta(t)$, we first combine
Eqs.~(\ref{1.44}) and (\ref{5.11}) to derive
(Appendix~\ref{app:2.7})
\begin{multline}
  \label{7.1}
    F(\omega, t)=
    \frac{d_{21}}{\sqrt{\pi \hbar \epsilon _0  \mathcal{A}}}
    \sqrt{\frac{c}{\omega}}
    \frac{\omega^2}{c^2}
\\[.5ex]
\times
      \int ^t _0 \D t'\,
     G^*(0^+, z_A, \omega)\tilde{C}_2^*(t') e^{i\omega(t-t')}
     e^{i
\tilde{
\omega
}
_0t'}
   .
\end{multline}
\begin{figure}[t]
\includegraphics[width=.9\linewidth]{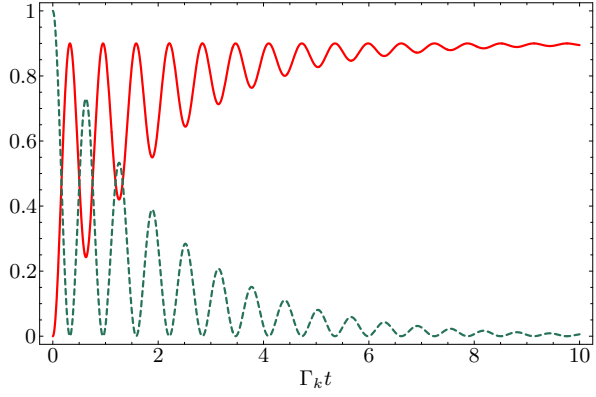}
\caption{\label{fig2}
The efficiency of one-photon
Fock-state preparation, $\eta(t)$,
Eq.~(\ref{7.5}), (solid curve) and the atomic
upper-state occupation probability
$|
\tilde{
C
}
_2(t)|^2, $ Eq.~(\ref{1.41}), (dashed curve) are shown for
$\gamma_{k\mathrm{rad}}$ $\!=$ $\!0.9$ $\!\Gamma_k$,
\mbox{$\omega_k$ $\!-$ $\!\tilde{\omega}_0$
$\!=$ $\!0.1$ $\!\Gamma_k$},
\mbox{$
R_k$
$\!=$ $\!10$ $\!\Gamma_k$},
\mbox{$\omega_k$ $\!=$ $\!2$ $\!\times$ $\!10^8$ $\!\Gamma_k$}.}
\end{figure}%
Next, we insert Eq.~(\ref{1.41}) into Eq.~(\ref{7.1}),
make use of the Green function as given by Eq.~(\ref{app.1}),
and perform the $t$~integration. Omitting off-resonant terms,
we obtain
\begin{multline}
  \label{7.3}
F(\omega, t)=
    \frac{\kappa _k} {2}
    \frac{1}
    {(\omega - {\tilde \omega}_0/2- \Omega_k^*/2 )^2 -\zeta_k^{*2}/4}
\\[.5ex]\times
\left\lbrace
  e^{i\omega t}
   \left[
     1
      - e^{-i(\omega - {\tilde \omega}_0) t}
      C_2^*(t)
   \right]
\right.
\\[.5ex]
\left.
   - \frac{i\alpha_k^* \Omega_k^*}{2 \zeta_k^*}
   \frac{\sin(\zeta_k^* t/2)}
   {\omega - \Omega _k^*}
   e^{i( {\tilde \omega}_0+\Omega_k^*) t/2}
   \right\rbrace,
\end{multline}
where
\begin{multline}
  \label{7.4}
    \kappa_k
    =
    -\omega_k
     \frac{d_{21}}{\sqrt{\pi \hbar \epsilon_0 A}}
    \sqrt{\frac{c}{\omega _k}}
\\
\times
    \frac{t_{13}^*(\Omega _k)\, e^{-i\omega_k n_1
(\Omega_k)
    l /c}}{|n_1
(\Omega_k)
    |^{2} l}
    \sin(\omega_k |n_1
(\Omega_k)
    | z_A/c).
\end{multline}

In what follows we assume that \mbox{$\gamma_k$ $\!<$
$\!\Gamma_k$}. Note, that in the opposite case
superstrong coupling can be observed
(see, e.g., Ref.~\cite{meiser:065801}).
Combining Eqs.~(\ref{5.19}) and (\ref{7.3})
and recalling Eq.~(\ref{1.41}), we
arrive, after some calculation, at
the following expression for the efficiency
($\gamma_k$ $\!<$ $\!\Gamma_k$):
\begin{equation}
  \label{7.5}
    \eta(t) = \frac{\gamma_{k\mathrm{rad}}
      \Gamma_k}{\Gamma _k^2-\gamma_k^2}
    \frac{R_k^2}{\rho_k^2+\Gamma_k^2/4}
    \left[
      1-|
\tilde{
C
}
_2(t)|^2
    \right],
\end{equation}
where
\begin{equation}
\label{7.6} \gamma_{k\mathrm{rad}}
     = \frac{c}{2 |n_1
(\Omega_k)
     | l}   |T_k|^2
\end{equation}
with
\begin{equation}
  \label{7.6.1}
T_k =
  \frac{ t_{13}(\Omega _k)} {
   \sqrt{|n_1
(\Omega_k)
     |}}
\, e^{i\omega_k n_1
(\Omega_k)
    l /c}
.
\end{equation}
In particular in the limit when $t\to\infty$, then
$\eta(t)\to[\gamma_{k\mathrm{rad}}\Gamma_k/(\Gamma_k^2-\gamma_k^2)]
[R_k^2/(\rho_k^2+\Gamma_k^2/4)]$, which approximately simplifies to
$\eta(t)\to\gamma_{k\mathrm{rad}}/\Gamma_k$ for a
sufficiently high-$Q$ cavity and almost exact resonance.
Note that this value is always observed
at the instants when the atom is in the lower state.
The behavior of the function $\eta(t)$ is illustrated in 
Fig.~\ref{fig2}. For comparison, the atomic upper-state
occupation probability $|\tilde{C}_2(t)|^2$ is also shown.


\subsubsection{Shape of the excited outgoing field}
\label{sec5.2.2}

To study the propagation of the excited outgoing
field in space and time, we consider
the operator of the electric field strength
\begin{equation}
  \label{7.41}
    \hat{E}^{(+)}_{\mathrm{out}}(z)
      = \int_0^{\infty}\!
      \D\omega\,
      e^{i\omega z/c}
      \left.
      \uh{E}_{\mathrm {out}} (z, \omega)
\right|_{z=0_+}
\end{equation}
($z$ $\!>$ $0$). Recalling Eq.~(\ref{5.3}), we may write
 \begin{equation}
  \label{7.43}
    \hat{E}^{(+)}_{\mathrm{out}}(z)
      =
      {\textstyle\frac{1}{2}}
      \int_0^{\infty}\!
      \D\omega\,
      \sqrt{\frac{\hbar\omega}{\varepsilon_0 c \pi\mathcal{A}}}\,
      e^{i\omega z/c}
  \hat{ b}_{\mathrm{out}}  (\omega)
\end{equation}
or equivalently, inserting Eq.~(\ref{5.13}) into Eq.~(\ref{7.43}),
\begin{equation}
  \label{7.45}
    \hat{E}^{(+)}_{\mathrm{out}}(z)
      =
          \sum_i
      \phi_i^*(z,t)
        \hat{b}_{\mathrm {out}\,i}(t)
,
\end{equation}
where
\begin{equation}
  \label{7.47}
\phi_i(z, t)
    =
    {\textstyle\frac{1}{2}}
     \int_0^{\infty}\!
     \D\omega\,
      \sqrt{\frac{\hbar\omega}{\varepsilon_0 c \pi\mathcal{A}
        }}\,
     e^{-i \omega z/c}
    F_i(\omega, t)
.
\end{equation}
The intensity of the outgoing field at position 
$z$ is then determined by
\begin{equation}
  \label{7.32}
  I(z,t)
  =
   \left
          \langle
           \psi(t)
         \right|
\hat{E}^{(-)}_{\mathrm{out}}(z)
\hat{E}^{(+)}_{\mathrm{out}}(z)
         \left|
       \psi(t)
        \right
        \rangle
\end{equation}
with $\hat{E}^{(+)}_{\mathrm{out}}(z)$ and
$\hat{E}^{(-)}_{\mathrm{out}}(z)$ $\!=$
$\![\hat{E}^{(+)}_{\mathrm{out}}(z)]^\dagger$
from Eq.~(\ref{7.45}). Using Eqs.~(\ref{5.15}),
(\ref{5.6}), (\ref{5.17}), and (\ref{7.7}), we derive
\begin{equation}
  \label{7.33}
   I(z,t)
  =
  \eta(t)
  |\phi_1(z,t)|^2
,
\end{equation}
which reveals that $\phi_1(z,t)$
represents the spatio-temporal shape of the
outgoing field associated with
the excited mode $F_1(\omega,t)$, and
$\eta(t)$ is nothing but the expectation value
\mbox{$\left \langle\psi(t)\right|$ 
$\!\hat{b}^{\dagger}_{\mathrm{out}\,1}(t)$ 
$\!\hat{b}_{\mathrm {out}\,1}(t)$
$\!\left|\psi(t)\right\rangle$}.

For simplicity, let us restrict our attention to 
the case where the cavity is not filled with medium
($n_1$ $\!\equiv$ $\!1$) and assume that the thickness 
of the fractionally transparent mirror is small compared with 
the cavity length. Then, as shown in Appendix~\ref{app:2.9},
$\phi_1(z,t)$ can be regarded as also describing
the part of the excited outgoing wave packet
that may be still inside the cavity, i.\,e., 
\mbox{$-l$ $\!\leq$ $\!z$ $\!<$ $\!0$}. Hence, we may write
\begin{align}
  \label{7.7}
\phi_1(z,t)
&
=
\Theta(z+l)
  {\textstyle\frac{1}{2}}
     \int
     _0^{\infty}\!
     \D\omega\,
     \sqrt{\frac{\hbar\omega}{\varepsilon_0 c \pi\mathcal{A}
  }}\,
     e^{-i \omega z/c}
     F
_1
     (\omega, t)
\nonumber\\&
\simeq
\Theta(z+l)
  {\textstyle\frac{1}{2}}
\sqrt{\frac{\hbar\omega_k}{\varepsilon_0 c \pi\mathcal{A}
  }}\,
     \int
     _0^{\infty}\!
     \D\omega\,
     e^{-i \omega z/c}
     F
_1
     (\omega, t)
,
\end{align}
which, by means of Eqs.~(\ref{5.17}) and (\ref{7.3}),
can be evaluated to (approximately) yield
\begin{multline}
  \label{7.9}
\phi_1(z,t)=
\Theta(-z)
  \Theta(z+l)
  \phi_1^<  (z,t)
\\[.5ex]
  -
  \Theta(z)\Theta(ct-z) \phi_1^>  (z,t)
  ,
\end{multline}
where
\begin{multline}
  \label{7.14}
  \phi_1^>  (z,t)
  =
        \sqrt{
  {\displaystyle
        \frac{\pi\hbar\omega_k}
        {\varepsilon_0 c\mathcal{A} \eta(t)}
      }}
   {\displaystyle
     \frac{\kappa _k
      }
        {\zeta_k^*}
}
        e^{i({\tilde \omega}_0 + \Omega_k^*) (t-z/c)/2}
\\
\times
        \sin[\zeta_k^*(t-z/c)/2]
\end{multline}
and
\begin{multline}
  \label{7.12}
  \phi_1^<  (z,t)
  =
           \sqrt{
     {\displaystyle
        \frac{\pi\hbar\omega_k}
        {\varepsilon_0 c\mathcal{A} \eta(t)}
      }}
   {\displaystyle
     \frac{\kappa _k
      }
        {\zeta_k^*}
}
        e^{i\Omega_k ^*(t-z/c)}
\\
\times
        e^{i({\tilde \omega}_0-\Omega_k^*) t/2}
        \sin(\zeta_k^*t/2).
\end{multline}
However note that in Eq.~(\ref{7.45}) describing
the outgoing field outside 
as well as
inside the cavity, the outside- and inside-parts of
$\phi_i(z,t)$ are in general associated with different 
operators $\hat{b}_{\mathrm {out}\,i}(t)$, because of 
absorption in the fractionally transparent mirror.

\begin{figure}[t]
\includegraphics[width=.9\linewidth]{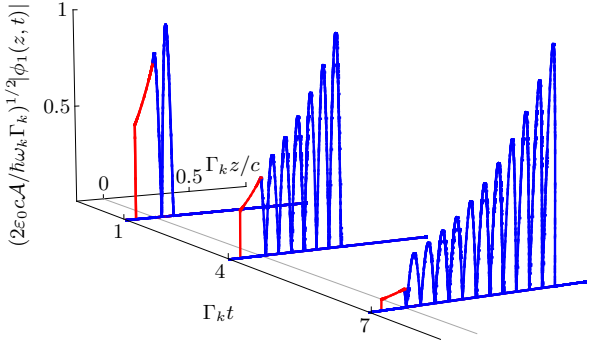}
\caption{\label{fig3}The spatio-temporal
behavior of the excited outgoing
wave packet,
$|\phi_1(z,t)|$, Eq.~(\ref{7.9}),
in the case of continuing atom--field
interaction for \mbox{$\Gamma _k l/c$ $\!=$ $\!0.7$}.
The parameters are the same as in Fig.~\ref{fig2}.
}
\end{figure}%
The behavior of the absolute value of
$\phi_1(z,t)$ as a function of $t$ and $z$
is illustrated in Fig.~\ref{fig3}.
Comparison with Fig.~\ref{fig2} reveals that
$|\phi_1^>(z,t)|$ oscillates according to the Rabi
frequency of the atom--field interaction.
It should be stressed that in Eq.~(\ref{5.21})
the argument $\beta_1(t)$ of the characteristic function
$C_1[\beta_1(t),t]$ of the quantum state of the
excited outgoing mode $F_1(\omega,t)$ refers to
the wave packet $\phi_1(z,t)$ as a whole, i.\,e., 
it refers not only to its part outside the cavity but
also to its part inside the cavity, which is observed 
for short times ($\Gamma_k t \lessapprox 1$)
when the emitted photon belongs to the cavity and 
the world outside the cavity simultaneously.


\subsection{Short-term atom--field interaction}
\label{sec5.3}

Let us now consider the case where the atom
leaves the cavity at some finite time $\tau$ so that
the interaction of the atom with the cavity-assisted field
effectively terminates at this time. Whereas for times $t$ in the
interval \mbox{$0$ $\!\leq$ $t$ $\!\leq$ $\!\tau$},
the state vector $|\psi(t)\rangle$ is again given by
Eq.~(\ref{1.17}), it reads
\begin{equation}
\label{7.11-0}
|\psi(t)\rangle = e^{-i(t-\tau)\hat{H}_0/\hbar}
|\psi(\tau)\rangle
\end{equation}
if \mbox{$t$ $\!\ge$ $\!\tau$}. Here,
$\hat{H}_0$ is the Hamiltonian of the uncoupled
system, i.\,e., the sum of the first two terms in Eq.~(\ref{1.15}),
and $|\psi(\tau)\rangle$ is given by $|\psi(t)\rangle$ from
Eq.~(\ref{1.17}) for $t$ $\!=$ $\!\tau$.
Hence, Eq.~(\ref{7.11-0}) can be written as
(\mbox{$t$ $\!\ge$ $\!\tau$})
\begin{multline}
\label{7.11}
  |\psi(t)\rangle =
  C_2(\tau)e^{-i\omega_{0} t}
  |2\rangle|\!\left\lbrace0\right\rbrace\!\rangle
\\[1ex]
   +\int \D z \int_0^\infty \D \omega\,
   C_1(z, \omega, \tau) e^{-i\omega t}
   |1\rangle\hat{f}^{\dagger}(z, \omega)
   |\!\left\lbrace0\right\rbrace\!\rangle.
\end{multline}
Note that the condition
\mbox{$\rho_k \tau$ $\!\gg $ $\!1$} is required in order
to observe damped vacuum Rabi oscillations.


\subsubsection{One-photon-Fock-state extraction efficiency}
\label{sec5.3.1}

To calculate the quantum state of the outgoing field
in the case of short-term atom-field interaction, we
insert Eq.~(\ref{7.11}) in Eq.~(\ref{5.7}) and
use Eq.~(\ref{5.3}) together with Eq.~(\ref{5.1}).
In this way, we again arrive at Eq.~(\ref{5.15-1}), 
but now with
\begin{multline}
  \label{7.13}
F(\omega, t,\tau)
=
    -2i
   \sqrt{\frac{c}{\omega}}
    \frac{\omega^2}{c^2}
\\[.5ex]\times
      \int \D z'\,
       \sqrt{\varepsilon''(z', \omega)}\,
     G^*_{\rm out}(0^+,z',\omega)
     C_1^*(z',\omega, \tau)
     e^{i\omega t}
\end{multline}
in place of $F(\omega,t)$. Comparing Eq.~(\ref{7.13}) 
with Eq.~(\ref{5.11}), we easily see that
\begin{align}
  \label{7.15}
F(\omega, t,\tau)=
   e^{i\omega (t- \tau)}
F(\omega, \tau) .
\end{align}
Choosing [$F_i(\omega,t)\mapsto F_i(\omega,t,\tau)$]
\begin{equation}
  \label{7.26}
    F_1(\omega,t,\tau) =
    \frac{F(\omega , t,\tau)}{\sqrt{\eta(t,\tau)}}\,,
\end{equation}
we are again left with an equation of the form of Eq.~(\ref{5.21})
[together with Eqs.~(\ref{5.21-2}) and (\ref{5.21-3})], where,
according to Eq.~(\ref{5.19}) [$\eta(t)\mapsto\eta(t,\tau)$],
the efficiency of preparation of the excited outgoing mode 
in a one-photon Fock state now reads
\begin{equation}
 \label{7.16}
\eta(t,\tau)  = \int
_0^{\infty}\!
         \D\omega\, |F(\omega ,t,\tau)| ^2
=\eta(\tau)
.
 \end{equation}
Hence, $\eta(t,\tau)$ does not depend on $t$;
it is simply given by the efficiency observed in the case 
of continuing atom--field interaction at time $\tau$, 
i.\,e., by $\eta(t)$ from Eq~(\ref{5.19}) for $t$ $\!=$ $\!\tau$.


\subsubsection{Shape of the excited outgoing field}
\label{sec5.3.2}

\begin{figure}[b]
\includegraphics[width=.9\linewidth]{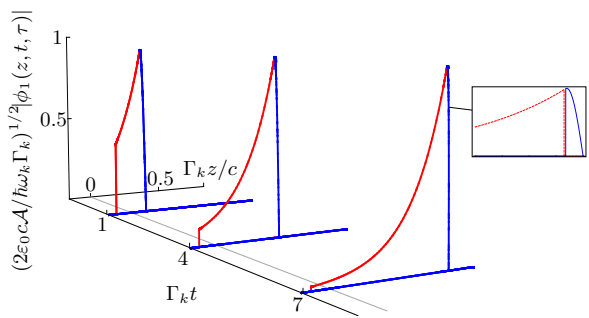}
\caption{\label{fig4}The spatio-temporal behavior
of the excited outgoing
wave packet
$|\phi_1(z,t,\tau)|$,
Eq.~(\ref{7.17}), in the case of short-term atom--field
interaction for \mbox{$\Gamma _k$ $\!\tau$}
$\!=$ $\!0.3$, \mbox{$\Gamma _k l/c$ $\!=$ $\!0.7$}.
The parameters are the same as in Fig.~\ref{fig2}.
The inset shows the leading edge (solid curve)
and the trailing edge (dashed curve).
}
\end{figure}%
A calculation in line with that leading
from Eq.~(\ref{7.41}) to Eq.~(\ref{7.9}), now yields 
the following form of the excited outgoing mode
[$\phi(z,t)\mapsto\phi(z,t,\tau)$]:
\begin{multline}
  \label{7.17}
  \phi_1(z,t,\tau)
=
    \Theta(ct-c\tau-z)
    \Theta(z+l)
   \phi_1^<  (z,t,\tau)
\\[.5ex]
  + \Theta(z-ct+c\tau)\Theta(ct-z)
  \phi_1^>  (z,t,\tau)
,
\end{multline}
where
\begin{multline}
  \label{7.20}
\phi_1^>  (z,t,\tau)
  =     \sqrt{
     {\displaystyle
        \frac{\pi\hbar\omega_k}
        {\varepsilon_0 c\mathcal{A} \eta(t)}
      }}
   {\displaystyle
     \frac{\kappa _k
      }
        {\zeta_k^*}\,
}
        e^{i( \tilde {\omega}_0 + \Omega_k^*) (t-z/c)/2}
\\
\times
        \sin[\zeta_k^*(t-z/c)/2]
\end{multline}
and
\begin{multline}
  \label{7.18}
\phi_1^<  (z,t,\tau)
  = \sqrt{
 {\displaystyle
        \frac{\pi\hbar\omega_k}
        {\varepsilon_0 c\mathcal{A} \eta(t)}
      }}
  {\displaystyle
    \frac{\kappa _k
     }
        {\zeta_k^*}\,
}
      e^{i\Omega_k ^*(t-z/c)}
\\
\times
        e^{i( \tilde {\omega}_0-\Omega_k^*) \tau/2}
        \sin(\zeta_k^*\tau/2).
\end{multline}
Note that Eq.~(\ref{7.20}) agrees with Eq.~(\ref{7.14}).
The behavior of the absolute value of $\phi_1(z,t,\tau)$
is illustrated in Fig.~\ref{fig4}, where the interaction
time $\tau$ is chosen in such a way that it
is the time at which the atom is the first time in the
lower state (cf.~the upper-state
occupation probability shown in Fig.~\ref{fig2}).

\begin{figure}[t]
\includegraphics[width=.9\linewidth]{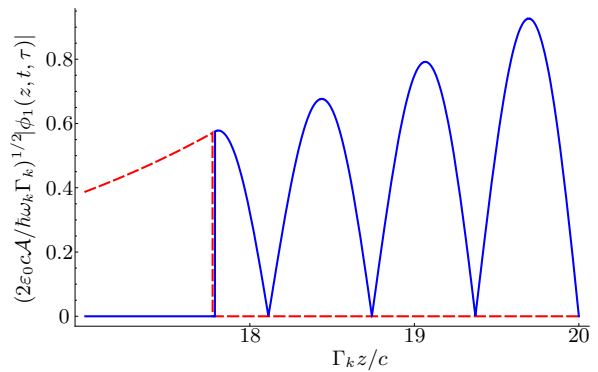}
\caption{\label{fig5}The spatio-temporal shape
of the excited outgoing mode,
$|\phi_1(z,t,\tau)|$, Eq.~(\ref{7.17}),
for $\Gamma_k \tau$ $\!=$ $\!2.2$,
$ \Gamma_k t$ $\!=$ $\!20$,
and \mbox{$\Gamma _k l/c$ $\!=$ $\!0.7$}.
The parameters are the same as in Fig.~\ref{fig2}.
The solid (dashed) curve shows the leading (trailing) edge.
}
\end{figure}%
We see that the excited outgoing field has,
for the chosen interaction times $\tau$,
the form of a single-peaked pulse,
whose trailing and leading edges are
determined by $\phi_1^<(z,t,\tau)$
and $\phi_1^>(z,t,\tau)$, respectively.
Since $\phi_1^>(z,t,\tau)$ approaches zero as $\tau$
tends to zero, the pulse can be regarded as being 
fully determined by $\phi_1^<(z,t,\tau)$
for sufficiently short interaction times.
The situation can drastically change when
longer interaction times are considered
so that $\tau$ cannot be regarded as being small
compared to $\Gamma_k^{-1}$. In this case,
the contribution to $\phi_1(z,t,\tau)$
of $\phi_1^>(z,t,\tau)$ can become
the dominating one, and, if the atom is allowed 
to undergo Rabi oscillations before it leaves the 
cavity, a multi-peaked pulse is observed,
as can be seen from Fig.~\ref{fig5}.


\subsection{Comparison with quantum noise theories}
\label{sec5.4}

Let us compare the results with the ones
obtained by using methods of QNT. As QNT is based on the
assumption that the electromagnetic fields inside and
outside the cavity represent independent degrees of freedom, 
which give rise to two separate Hilbert spaces, there is 
in particular no way to consider (excited) outgoing
modes that simultaneously belong to the cavity
and the outside world. To go into more details,
let us first consider the regime of continuing
atom--field interaction.


\subsubsection{Continuing atom--field interaction}
\label{sec5.4.1}

To describe the 
excited outgoing
field inside and outside the cavity
from the point of view of QNT, one could consider 
wave packets of the types of $\phi_1^>(z,t)$ [Eq.~(\ref{7.14})] 
and $\phi_1^<(z,t)$ [Eq.~(\ref{7.12})], respectively,
and introduce functions $F^{>}_1(\omega,t)$ and 
$F^{<}_1(\omega, t)$ according to
\begin{equation}
  \label{7.28}
  F
  ^{>(<)}
  _1
      (\omega, t)
    =
    \frac{1}{\sqrt{2\pi c \mathcal{N}_1^{>(<)}(t)}}
    \int
    _{>(<)}\!\!
     \D z\, e^{i \omega z/c}
     \phi_1^{>(<)}(z,t)
     ,
\end{equation}
where
\begin{equation}
  \label{7.25}
\mathcal{N}_1
  ^{>(<)}(t) =  \int_{>(<)}
  \!
  \D z\,
  \bigl|\phi_1^{>(<)}  (z,t)\bigr| ^2.
\end{equation}
Here, the integral $\int_{>(<)}\! \D z\ldots$ runs over
the interval \mbox{$0$ $\!<$ $\!z$ $\!<$ $\!ct$}
(\mbox{$-l$ $\!<$ $\!z$ $\!<$ $\!0$}). Now, one could 
identify $F_1(\omega,t)$ in Eq.~(\ref{5.15-1}) 
for $C_\mathrm{out}[\beta_i(t),t]$ with $F_1^>(\omega,t)$.
Since for the field inside the cavity, Eq.~(\ref{5.15-1})
must be replaced by Eq.~(\ref{app2.48}), one could also 
identify $F_1(\omega,t)$ with $F_1^<(\omega,t)$.
Disregarding the `interference' terms, which prevent
$C_\mathrm{out}[\beta_i(t),t]$ from being a product,
one may introduce the single-mode characteristic function
\begin{equation}
\label{7.29}
C_1^{>(<)}(\beta,t) =
e^{-|\beta|^2/2}
   \left[1-\eta^{>(<)}(t)
   |\beta|^2\right]
\end{equation}
together with
\begin{equation}
 \label{7.30}
\eta^{>(<)}(t) =
\left|\int_0^\infty\D\omega\,F_1^{>(<)}(\omega,t)
F^{>(<)*}(\omega ,t)\right|^2,
\end{equation}
where, according to Eqs.~(\ref{5.15-1}) and 
(\ref{app2.48}), respectively,
\begin{equation}
  \label{7.35}
  F^{>}(\omega ,t)
  =
  F(\omega ,t)
\end{equation}
and
\begin{equation}
  \label{7.36}
  F^{<}(\omega ,t)
  = \sqrt{\frac{\Gamma_k} {\gamma_{k\mathrm{rad}}}}\,
  F(\omega ,t).
\end{equation}
Obviously, Eq.~(\ref{7.29}) together with Eq.~(\ref{7.30})
replaces Eq.~(\ref{5.21-2}) together with Eq.~(\ref{5.19}).
Using Eqs.~(\ref{5.17}), (\ref{7.7}), (\ref{7.28}),
(\ref{7.28}), and (\ref{7.25}), after some calculations
we find that $\eta^{>(<)}(t)$ can be written as
\begin{align}
 \label{7.31}
& \eta^{>}(t) =
 \frac{2\varepsilon_0 \mathcal{A}}
        {\hbar\omega_k }
\eta(t)\mathcal{N}_1^{>}(t),
\\[1ex]
 \label{7.37}
&\eta^{<}(t) = \frac{\Gamma_k} {\gamma_{k\mathrm{rad}}}\,
 \frac{2\varepsilon_0 \mathcal{A}}
        {\hbar\omega_k }
  \eta(t)\mathcal{N}_1^{<}(t),
\end{align}
with $\eta(t)$ from Eq.~(\ref{5.19}).

\begin{figure}[t]
\includegraphics[width=.9\linewidth]{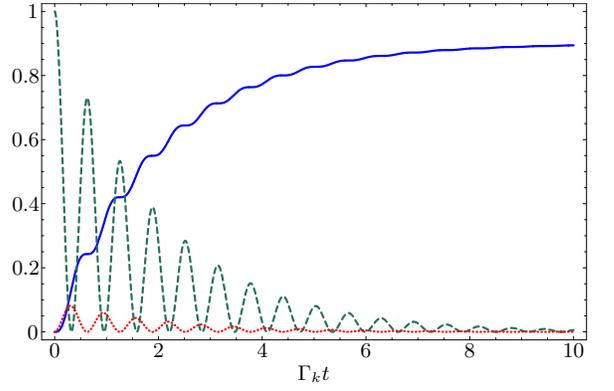}
\caption{\label{fig9}
The quantities $\eta^>(t)$ (solid curve)
and $\eta^<(t)$ (dotted curve),
c.f. Eqs.~(\ref{7.31}) and (\ref{7.37})
(\mbox{$\Gamma _k l/c$ $\!=$ $\!0.7$}),
and the atomic upper-state occupation probability
$|\tilde{C}_2(t)|^2, $ Eq.~(\ref{1.41}), (dashed curve).
The parameters are the same as in Fig.~\ref{fig2}.
}
\end{figure}%
To give a physical explanation of Eqs.~(\ref{7.31}) 
and (\ref{7.37}), we first note that from
Eq.~(\ref{7.33}) it follows that
\begin{equation}
  \label{7.34}
   I^{>}
(t)
  =
  \int_{>}
  \!
  \D z\,
   I(z,t)
=
 \eta^{>}(t)
.
\end{equation}
From the theory of photodetection of light \cite{vogel} 
we know that the probability of registering a photon
during the time interval $[0,t]$ by a detector at $z$ $\!=$
$\!0_+$ is proportional to $I^>(t)$. Hence, the quantity
$\eta^>(t)$, which refers to the part of the excited outgoing
wave packet that is quite outside the cavity,
cannot be regarded, in general, as being the
efficiency of preparation of the outgoing field in a 
one-photon Fock-state; it can be merely regarded as being
proportional to the probability of registering an emitted 
photon outside the cavity during the chosen time interval.
To clarify the meaning of the quantity $\eta^<(t)$, 
which refers to the part of the excited outgoing
wave packet that is quite inside the cavity,
we recall that the total excited field inside the 
cavity consists of a part traveling from the left 
to the right and a part traveling from the right 
to the left. Obviously, $\eta^<(t)$  
refers to the (small) fraction of the former part 
which is transmitted through the fractionally
transparent  mirror. 
It is hence proportional to the
probability of registering a photon if this fraction
of the part of the excited field inside the cavity
which travels from the left to the right could be detected.  
The dependence on $t$ of $\eta^{>}(t)$ and $\eta^{<}(t)$
is illustrated in Fig.~\ref{fig9}. As we can see,
$\eta^{>}(t)$ almost monotonically increases with time,
while $\eta^{<}(t)$ decreases with increasing time
in an oscillatory manner. As expected, $\eta^{>}(t)$ 
approaches $\eta(t)$ as $\Gamma_k t$ tends to infinity 
(cf.~Fig.~\ref{fig2}). Clearly, in the limit when  
$\Gamma_k t\to\infty$, the wave packet $\phi_1(z,t)$ 
associated with the excited outgoing mode $F_1(\omega,t)$
is strictly localized outside the cavity and 
$C_1^{>}(\beta,t)$ equals
the characteristic function $C_1(\beta,t)$ of
the quantum state of the excited outgoing mode.

Since the wave packet associated with the excited 
outgoing mode covers the areas inside and outside the cavity
in general, a photon carried by the mode belongs 
simultaneously to the two areas in general.
To model this effect within the framework of QNT,
where the fields inside and outside the cavity are 
considered as 
belonging to
two different
Hilbert spaces,
one had to introduce (on disregarding absorption losses)
entangled states between a photon inside the cavity 
and a photon outside the cavity. Needless to say that 
from the point of view of QED, such a concept would be
rather artificial.


\subsubsection{Short-term atom--field interaction}
\label{sec5.4.2}

Let us compare the results
with the ones in Ref.~\cite{khanbekyan:053813},
where---in analogy to QNT---cavity modes are explicitly
introduced and, on this basis, input--output relations are derived.
In particular, let us consider a single excited cavity mode,
say the $k$th mode, and assume that at initial time 
\mbox{$t$ $\!=$ $\!0$} this mode is prepared in an excited 
state and the field outside the cavity is in the vacuum state. 
This means that the time of preparation of the mode in the
excited state (which corresponds to the time $\tau$ of 
the atom--field interaction in Sec.~\ref{sec5.3}) must be
much smaller than its decay time $\Gamma_k^{-1}$.
In the frequency interval $\Delta_k$, the operator input--output
relation in the Heisenberg picture can then be written
as \cite{khanbekyan:053813}
\begin{equation}
\label{9.31}
        \hat{b}_{k \mathrm {out}}
        (\omega, t) =
        F^*_{k\mathrm{rel}} (\omega, t)
        \hat{a}_k(0)
        + \hat{B}_k(\omega,t),
\end{equation}
where $\hat{a}(0)$ is the photon destruction operator of the
cavity mode at the initial time, $\hat{B}_k(\omega,t)$ is a 
linear functional of the operators of the input field
$\hat{b}_{k\rm{in}}(\omega,0)$ and the noise sources 
$\hat{c} _{k\lambda}(\omega,0)$ both taken at the initial
time (see Appendix \ref{app:3}), and
\begin{align}
\label{9.33}
        F_{k\mathrm{rel}} (\omega, t)
&=
         \frac{i} {\sqrt{2\pi}}
        \left[\frac{c}{2n_1(
\Omega_k
) l}\right]^{\frac{1}{2}}
        \!
        T_k^*
        \,
        e^{i\omega t}
\nonumber\\[.5ex]
&\quad
\times \, \frac
         {\exp \left[-i (\omega - \Omega ^* _k) (t+\Delta t)
                \right] -1
         }
         {\omega - \Omega ^* _k}
\end{align}
\mbox{($\Delta t$ $\!\to$ $\!0_+$)}.
From Eq.~(\ref{9.31}) one can conclude that
$F_{k\mathrm{rel}}(\omega, t)$ represents the part
of the output field relevant to the excited
cavity mode. The efficiency to find at time~$t$ this
part---referred to as the relevant (nonmonochromatic) 
output mode---in a one-photon Fock state when the cavity 
mode has initially been prepared in a one-photon Fock 
state then reads as
\cite{khanbekyan:053813}
\begin{align}
        \label{9.32}
        \eta
        _{k\mathrm{rel}} (t)
&=
         \int_0^\infty \D\omega\, |F_{k\mathrm{rel}}(\omega ,t)| ^2
\nonumber\\[0.5ex]
        &
\simeq
        \frac{
        \gamma_{k\mathrm{rad}}
        }{\Gamma _k}
        \left[
     1-e^{-\Gamma _kt}
   \right]
,
\end{align}
which is seen to be time dependent, in contrast to
the time-independent efficiency (\ref{7.16}).

This result is not surprising, since---in contrast to the
relevant mode, which by construction defines a wave packet 
entirely outside the cavity---the outgoing wave packet
$\phi_1(z,t,\tau)$ [Eq.~(\ref{7.17})], which
corresponds to the excited mode that really carries the 
photon, covers simultaneously the areas inside and outside 
the cavity (cf.~Fig.~\ref{fig4}). As in the case of 
continuing atom--field interaction, it is natural to 
expect that only in the case when the condition 
\mbox{$\Gamma_k t$ $\!\gg$ $\!1$} holds, then 
$\eta_{k\mathrm{rel}}(t)\simeq\eta(\tau)$ for a 
value of $\tau$ for which the atom is in the ground state.

Returning to Sec.~\ref{sec5.3},
let us now 
consider the leading edge of $\phi_1(z,t,\tau)$ and
that part of the trailing edge of $\phi_1(z,t,\tau)$
which is entirely outside the cavity, and, in
line with Eq.~(\ref{7.28}), introduce the functions
($t\ge\tau$)
\begin{multline}
  \label{9.45}
  F^{>(<)}_{1}(\omega, t,\tau)
    =
    \frac{1}{\sqrt{2\pi c \mathcal{N}^{>(<)}_{1}(t,\tau)}}
\\[.5ex]\times
    \int_{>(<)}\!
     \D z\, e^{i \omega z/c}
     \phi_1^{>(<)}(z,t,\tau)
     ,
\end{multline}
where $\int_{>(<)}\! \D z\ldots$ runs over
the interval \mbox{$c(t-\tau)$ $\!<$ $\!z$ $\!<$ $\!ct$}
[\mbox{$0$ $\!<$ $\!z$ $\!<$ $\!c(t-\tau)$}], and
\begin{equation}
  \label{9.46}
\mathcal{N}_{1}^{>(<)}
  (t,\tau) =  \int_{>(<)}
  \!\D z\,
  \bigl|\phi_1^{>(<)}  (z,t,\tau)\bigr| ^2
,
\end{equation}
so that, in view of QNT, a characteristic function 
of the form of Eq.~(\ref{7.29}) could be defined, with
$\eta^{>(<)}(t)$ being replaced by
\begin{equation}
 \label{9.47}
 \eta^{>(<)}(t,\tau) =
 \left|\int_0^\infty \D\omega\,
   F^{>(<)} _{1}(\omega, t,\tau)
   F^*(\omega ,t,\tau)\right|^2.
\end{equation}
Straightforward calculation,
using Eqs.~(\ref{9.45}) and (\ref{9.46}) together with
Eqs.~(\ref{7.7}), (\ref{7.26}), and (\ref{7.17}), yields
\begin{equation}
        \label{9.34}
  \eta^{>(<)}(t,\tau)
       =
        \frac{2\varepsilon_0 \mathcal{A}}
        {\hbar\omega_k }
        \eta(\tau)
        \mathcal{N}_{1}^{>(<)}(t,\tau).
\end{equation}
\begin{figure}[t]
\includegraphics[width=.9\linewidth]{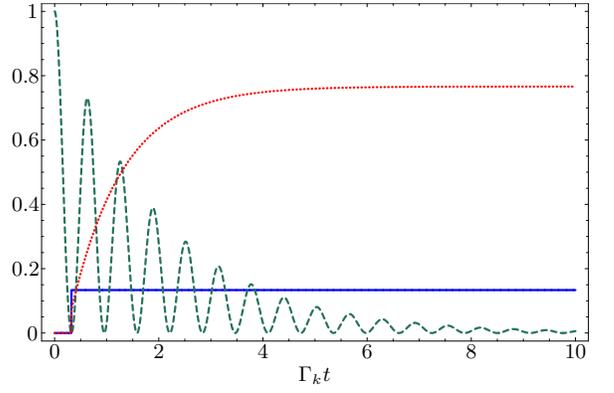}
\caption{\label{fig8}
The quantities $\eta^>(t,\tau)$ (solid curve) 
and $\eta^<(t,\tau)$ (dotted curve), Eq.~(\ref{9.38}),
and the atomic upper-state occupation probability
$|\tilde{C}_2(t)|^2, $ Eq.~(\ref{1.41}), (dashed curve)
for \mbox{$\Gamma_k\tau$ $\!=$ $\!0.3$},
\mbox{$\Gamma _k l/c$ $\!=$ $\!0.7$}.
The parameters are the same as in Fig.~\ref{fig2}.
}
\end{figure}%
In particular,
\begin{equation}
 \label{9.38}
 \eta^{<}(t,\tau)
=
\frac{R_k^2}{|\zeta _k|^2}
\left| \sin (\zeta _k \tau /2) \right| ^2 e^{-\Gamma _k \tau/2}
  \eta _{k\mathrm{rel}} (t-\tau)
,
\end{equation}
with $\eta _{k\mathrm{rel}}(t)$ according to Eq.~(\ref{9.32}).
The behavior of $\eta^{>}(t,\tau)$ and $\eta^{<}(t,\tau)$ 
as functions of~$t$ is illustrated in Fig.~\ref{fig8}.
To make contact with the results in Ref.~\cite{khanbekyan:053813},
we assume strong atom--field coupling, almost exact resonance,
$R_k^2/|\zeta_k|^2\simeq 1$, and short interaction time, 
such that \mbox{$\Gamma_k\tau\ll 1$} as well as
\mbox{$| \sin (\zeta _k \tau /2)| ^2\simeq 1$}, i.\,e.,
\mbox{$|\tilde{C}_2(\tau)|^2\simeq 0$}.
In this limiting case we have
\mbox{$\eta^{<}(t,\tau)\simeq\eta_{k\mathrm{rel}}(t)$}.
Recalling, that when \mbox{$\Gamma_k\tau\ll 1$}, then
the contribution of the leading edge $\phi_1^{>}(t,\tau)$
to $\phi_1(t,\tau)$ can be neglected, 
$\phi_1^{<}(t,\tau)\simeq\phi_1(t,\tau)$,
we conclude that $\eta_{k\mathrm{rel}}(t)\simeq\eta(\tau)$
if $\Gamma_k t$ $\!\gg$ $\!1$.
From reasons similar to those applied to the
case of continuing atom--field interaction it follows that
$\eta^{<}(t,\tau)\simeq\eta_{k\mathrm{rel}}(t)$ cannot be
regarded, in general, as being the efficiency of preparation
of the outgoing field in a single-photon Fock-state; it
may be merely regarded as being the probability of detection
of a photon outside the cavity.

Finally, let us address the following point. In QNT it is 
commonly assumed that (at equal times) input-field and 
cavity-field variables commute. On the other hand, the
QED approach in Ref.~\cite{khanbekyan:053813} shows that
\begin{equation}
  \label{9.36}
  \left[
     \hat{a}_k,
     \hat{b}_{k{\rm in}}^{\dagger}  (\omega)
   \right]
         =
F_{k} (\omega),
\end{equation}
where
\begin{equation}
  \label{9.71}
F_{k} (\omega)
=
\sqrt{\frac{c}{2 l}}
 \sqrt{\frac{\Gamma_k}
           {\gamma_{k\mathrm{rad}}}
       }
              \frac{T_k}
             {\sqrt{2 \pi}}
             \frac {i}
             {\omega - \Omega _k}
.
\end{equation}
Hence the question rises of whether or not the commutator
in Eq.~(\ref{9.36}) can be effectively set equal to zero.

To give an answer, we introduce the nonmonochromatic 
input mode that is related to this commutator,
\begin{align}
  \label{9.73}
    \hat{b}_{k\mathrm{in}}(t)
    =\int_0^\infty \D \omega\,
    F_{k\mathrm{in}} (\omega, t)
    \hat{b}_{k\mathrm{in}}(\omega)
,
\end{align}
where
\begin{equation}
\label{9.73-1}
F_{k\mathrm{in}}(\omega,t) = F_{k}(\omega) e^{-i\omega t}.
\end{equation}
Using Eq.~(\ref{9.31}) together with the formulas in 
Appendix~\ref{app:3}, one can calculate the corresponding 
nonmonochromatic mode of the output field,
\begin{equation}
  \label{9.37}
  F_{k \mathrm{out}} (\omega, t)
= \frac{r_{31}^\ast(\omega)}{|r_{31}(\omega)|}
\,F_k(\omega)e^{i\omega t}
,
\end{equation}
and show that
\begin{multline}
  \label{9.39}
  \int_0^\infty
  \D\omega\,
  F_{k\mathrm{rel}} ^*(\omega, t)
  F_{k \mathrm{out}} (\omega, t)
\\[.5ex]
  \simeq
  \int_{-\infty}^\infty\!
  \D\omega\,
  F_{k\mathrm{rel}} ^*(\omega, t)
  F_{k \mathrm{out}} (\omega, t)
= 0
.
\end{multline}
Hence, the relevant output mode as defined by Eq.~(\ref{9.33})
is not related to the commutator in question.
Applying Eq.~(\ref{7.7}) to the functions
$F_{k\mathrm{rel}}(\omega, t)$, Eq.~(\ref{9.33}), and
$F_{k \mathrm{out}} (\omega, t)$, Eq.~(\ref{9.37}),
one can calculate the corresponding spatio-temporal shapes
\begin{equation}
\label{9.41}
          \phi_{k\mathrm{rel}}(z,t)
          = -
\Theta (z)
\Theta (ct-z)\,
    {\textstyle\frac{1}{2}}
    \sqrt{\frac{\hbar\omega_k}{\varepsilon_0 l\mathcal{A}
  }}\,
        T_k^{*} \,
        e^{i\Omega_k^* (t-z/c)}
\end{equation}
and
\begin{multline}
  \label{9.43}
  \phi_{k\mathrm{com}}
  (z,t)
  =
  \Theta (z-ct)
\\[.5ex]
\times
 {\textstyle\frac{1}{2}}
    \sqrt{\frac{\hbar\omega_k}{\varepsilon_0 l\mathcal{A}
  }}\,
  \sqrt{\frac{\Gamma_k}
           {\gamma_{k\mathrm{rad}}}
       }
 \frac{r_{31}^*(\Omega _k)} {|r_{31}(\Omega _k)|}
  T_k
  e^{i\Omega_k (t-z/c)}
 ,
\end{multline}
respectively. Comparing Eqs.~(\ref{9.41}) and (\ref{9.43}) 
with Eq.~(\ref{7.17}), we see that, as expected,
$\phi_{k\mathrm{rel}}(z,t)$ matches, for $\Gamma_k\tau\ll 1$, 
that part of $\phi_1(z,t,\tau)\simeq\phi_1^<(z,t,\tau)$
which is entirely outside the cavity, whereas
$\phi_{k\mathrm{com}}(z,t)$ does not contribute to 
$\phi_1(z,t,\tau)$. We are thus left with the result 
that when the interaction time is sufficiently short,
then the commutator in question can be effectively
set equal to zero in the scheme under consideration.


\section{Summary and concluding remarks}
\label{summary}

Within the frame of macroscopic QED in dispersing and
absorbing media, we have given
an exact description of the resonant interaction of
a two-level atom in a \mbox{high-$Q$} cavity with
the cavity-assisted electromagnetic field,
with the atom and the field being initially
in the upper state and the ground state, respectively.
Using a source-quantity representation
of the electromagnetic field, we have performed
the calculations for a one-dimensional
cavity bounded by a perfectly
reflecting and a fractionally transparent mirror,
without regarding the fields inside
and outside the cavity as representing independent
degrees of freedom. We have applied the
theory to the determination of
(i) the Wigner function of the quantum state of the
excited outgoing mode and (ii) the spatio-temporal
shape of the wave packet 
corresponding to 
this mode.

As expected, the quantum state of the excited outgoing mode
is always a mixture of a one-photon Fock state and the
vacuum state, because of the unavoidable unwanted losses.
In the case of continuing atom--field interaction,
the efficiency of the mode being prepared in a
one-photon Fock state is time dependent and
features Rabi oscillation in the regime of strong
atom--field coupling. This is due to the fact
that, for not too long times, the 
corresponding
wave packet 
covers the areas both inside and outside the cavity
so that a photon emitted 
by the atom
belongs simultaneously to the two areas and can
thus be reabsorbed by the atom.
In particular, from the part of the 
wave packet
that is entirely localized outside the cavity, the
probability of registering the emitted photon by
a photodetector placed outside the cavity can be 
calculated---a quantity which, as expected, monotonously 
increases with time and approaches the efficiency 
of one-photon Fock-state preparation in the long-time limit.

In the case of short-term atom--field interaction,
where the atom leaves the cavity at some interaction
time~$\tau$, the efficiency of the excited outgoing mode
being prepared in a one-photon Fock state
is constant for all times \mbox{$t$ $\!\ge$ $\!\tau$}.
It is simply given by the value of the efficiency
observed at time $t$ $\!=$ $\!\tau$
in the case of continuing atom--field interaction.
This result again reflects the fact that
the field associated with the excited
outgoing mode covers the areas both inside
and outside the cavity. In particular,
if the atom leaves the cavity when it is
the first time in the ground state, then
this field can be regarded as a single-peak
pulse, the trailing edge of which covers the two areas.
Moreover, when the interaction time 
is sufficiently short, then 
the pulse
can be approximated by its
trailing edge, whose part that is entirely
localized outside the cavity matches the
field determined by the relevant mode appearing 
in the input--output relations derived 
within the frame of QNT.
Again, the part of the excited outgoing field
that is entirely localized outside the cavity
only determines the probability of registering the emitted 
photon by a photodetector placed outside the cavity 
and not the  efficiency of one-photon Fock-state preparation
in general.

The results particularly imply that the efficiency of preparing
the outgoing field in a one-photon Fock state, as
suggested from QNT, is in fact proportional to the probability
of registering a photon outside the cavity.
Only in the long-time limit the two quantities
become identical. As in QNT or related approaches
to cavity QED modes that simultaneously cover the
areas inside and outside a cavity are a priori
excluded from consideration, effects associated with
modes of this type cannot be exactly described by
such methods. What one could do to is to model them by 
introducing---somewhat artificially from the point of
view of QED---entangled states where a photon 
inside the cavity is entangled
with a photon outside the cavity.

The results also show that in the
short-time limit when the leading edge of the
excited outgoing field can be disregarded, then
the commutator in Ref.~\cite{khanbekyan:053813}
between the input field and the cavity field at equal
times can be effectively set zero, in agreement
with the QNT postulate. As we can see from Fig.~\ref{fig5},
the contribution to the outgoing mode which
corresponds to the leading edge increases with the 
atom--field interaction time. In practice this time
can be comparable with the time of extraction
of the state from the cavity (see, e.\,g.,
the scheme in Ref.~\cite{kuhn:067901})---a case
which with respect to more complicated
schemes, which may contain pumped multi-level atoms and/or
excited input fields,
requires further studies.


\begin{acknowledgments}
This work was supported by the Deutsche Forschungsgemeinschaft.
C.D.F. thanks Thomas Richter for useful discussions.
\end{acknowledgments}

\appendix


\section{Multilayer planar structure}
\label{app:1}

The one-dimensional cavity is modeled by a planar multilayer system,
where the layers \mbox{$j$ $\!=$ $\!0$}
and \mbox{$j$ $\!=$ $\!2$}, respectively, are assumed to
correspond to perfectly and fractionally reflecting mirrors, which
confine the cavity whose interior space corresponds to the
layer \mbox{$j$ $\!=$ $\!1$}.
With respect to the cavity axis $z$, we use shifted
coordinates such that
\mbox{$0$ $\!<$ $\!z$ $\!<$ $\!l$} for \mbox{$j$ $\!=$ $\!1$},
\mbox{$0$ $\!<$ $\!z$ $\!<$ $\!d$} for \mbox{$j$ $\!=$ $\!2$}, and
\mbox{$0$ $\!<$ $\!z$ $\!<\infty$} for \mbox{$j$ $\!=$ $\!3$}. The
(one-dimensional) Green function in the frequency domain reads
\cite{tomas:052103}
\begin{multline}
      \label{app.1}
\hspace{-2ex}
     G^{(jj')}(z, z', \omega )
      =
      {\textstyle\frac{1}{2}} i \bigl[
      \mathcal{E} ^{(j)>}    (z, \omega )\,
      \Xi^{jj'}    \mathcal{E} ^{(j')<}    (z', \omega)
      \Theta (j-j')
\\
      + \mathcal{E}^{(j)<}    (z, \omega )\,
      \Xi^{j'j}   \mathcal{E} ^{(j')>}   (z', \omega )
      \Theta (j'-j)
      \bigr] ,
\end{multline}
where the functions
\begin{equation}
      \label{app.3}
 \mathcal{E} ^{(j)>}    (z, \omega ) =
        e^{i \beta _j(\omega) (z-d _j)}
       + r  _{j/3}(\omega)
       e^{-i \beta _j(\omega) (z-d _j)}
      \end{equation}
and
\begin{equation}
      \label{app.5}
      {\mathcal{E}} ^{(j)<}    (z,   \omega )\, =\,
        e^{-i \beta _j(\omega) z}
      + r  _{j/0}(\omega)   e^{i \beta _j(\omega) z},
\end{equation}
respectively, represent waves of unit strength traveling
rightward and leftward in the $j$th layer and being reflected at
the boundary [note that $\Theta (j$ $\!-$ $\!j')$  means
\mbox{$\Theta (z$ $\!-$ $\!z')$} for $j$ $\!=$ $\!j'$]. Further,
$\Xi^{jj'}$ is defined by
\begin{equation}
      \label{app.7}
      \Xi^{jj'} =
      \frac{1}{\beta_3(\omega) t _{0/3}(\omega)}\,
      \frac{t _{0/j}(\omega)e^{ i \beta _j(\omega) d _j}}
      {D _{ j}(\omega)}\,
      \frac{t _{3/j'}(\omega)e^{ i \beta _{j'}(\omega) d _{j'}}}
      {D _{ j'}(\omega)}\,,
      \end{equation}
where
\begin{equation}
      \label{app.9}
      D_{j}(\omega) = 1 - r _{j/0}(\omega)
      r _{j/3}(\omega) e^{2 i \beta _j(\omega) d _j}
      \end{equation}
and
\begin{equation}
      \label{app.11}
      \beta _j(\omega) =
      \sqrt {\varepsilon _j (\omega)} \,\frac {\omega} {c}
      = [n_j ' (\omega) + i n_j '' (\omega) ]\,\frac {\omega} {c}
      \end{equation}
($d_1$ $\!=$ $\!l$, $d_2$ $\!=$ $\!d$, $d_3$ $\!=$ $\!0$).
The quantities \mbox{$t_{j/j'}(\omega)$ $\!=$
$\![\beta_j(\omega)/\beta_{j'}(\omega)]t_{j'/j}(\omega)$} and
$r_{j/j'}(\omega)$
denote, respectively, the
transmission and reflection coefficients between
the layers $j'$ and~$j$, which can be recursively determined.
Note that the zeros
of the function $D_1(\omega)$ [Eq.~(\ref{app.9}) for $j=1$
and $r_{10}$ $\!=$ $\!-1$]
determine the (complex) resonance frequencies $\Omega_k$
of the cavity under consideration,
\begin{equation}
      \label{app.13}
 D_1 (\Omega _{k})
    = 1+ r_{13}(\Omega _{k})
    e^{2 i \beta_1 (\Omega _{k}) l} = 0
.
\end{equation}
In particular, the part of the Green function
that relates the outgoing field at $z$ to the sources
at $z'$ in the $j'$th layer is given by
\begin{align}
  \label{app2.15}
  G_{\rm out}(z,z',\omega)
&  \equiv
  G^{(3j')}_{\rm out}(z, z', \omega )
\nonumber\\
& =
  G^{(3j')}(z, z', \omega )-
  G^{(3j')}_{\rm in}(z, z', \omega ),
\end{align}
where
\begin{align}
  \label{app2.17}
    G^{(3j')}_{\rm in}(z, z', \omega )
    =
        \frac{1}{2\beta_3(\omega)}
    i
     e^{-i \beta _3(\omega)z}
      e^{i \beta _{j'}(\omega)z'}
      \Theta(z'-z).
\end{align}


\section{Derivation of Eq.~(\ref{1.43})}
\label{app:2}

We decompose the Green tensor inside the cavity,
$G(z,z',\omega)\equiv
G^{(11)}(z,z',\omega)$, as given by Eq.~(\ref{app.1}),
into bulk and scattering
parts $G^0$ and $G^S$, respectively,
\begin{equation}
  \label{app2.1}
  G(z, z', \omega )
  =
  G^0 (z, z ', \omega )
  + G^S (z, z', \omega ),
\end{equation}
where
\begin{multline}
  \label{app2.3}
  G^0 (z, z ', \omega )
   =
    {\textstyle\frac{1}{2}} i \bigl[
    e^{i\beta_1 (\omega)(z-z')} \Theta(z-z')
\\
   +
    e^{-i\beta_1 (\omega)(z-z')} \Theta(z'-z)
\bigr],
\end{multline}
and
\begin{equation}
  \label{app2.5}
  G^S (z, z ', \omega )
   =\frac{g(z, z', \omega)}
   {D_1(\omega)}\,,
\end{equation}
with
\begin{multline}
  \label{app2.6}
  g (z, z ', \omega )
   =
    {\textstyle\frac{1}{2}}
    \frac{e^{i\beta_1 (\omega)l}}{\beta_1(\omega)} i
  \bigl[
    r_{13}(\omega)e^{-i\beta_1(\omega) (z-l)}
    {\mathcal{E}} ^{(1)<}    (z',   \omega )
\\
    -e^{i\beta_1(\omega) z}
    {\mathcal{E}} ^{(1)>}    (z',   \omega )
    \bigr] .
\end{multline}
Employing the residue theorem, we may Fourier transform
$\omega ^2 G^S (z, z ', \omega ) $ to obtain
\begin{align}
  \label{app2.7}
f(z, z', t)=&
  \int\!\frac{\mathrm{d}\omega}{2\pi}\,
  \omega ^2
  e^{-i\omega t}\,
\frac{g(z, z', \omega)}{D_1(\omega)}
\nonumber \\[1ex]
=&
\sum _k \frac{c}{2 n_1
(\Omega_k)
   l}\,\Theta (t)
   \Omega _k^2
   e^{-i\Omega _{k} t} g (z, z', \Omega _{k}).
\end{align}
Further, the inverse transformation reads
\begin{align}
  \label{app2.9}
 \omega^2  G^S (z, z ', \omega )
   =&
   \int \D t\,e^{i\omega t}
f(z, z', t)
   \nonumber \\[1ex]
   =&
   \sum _k \frac{c}{2 n_1
(\Omega_k)
       l}\,
   \frac{i\Omega _k^2}{\omega - \Omega_k}\,
    g (z, z', \Omega _{k}) .
\end{align}
Then, the integration in Eq.~(\ref{1.36}) can be
approximated by a principal value integration as
\begin{equation}
  \label{app2.8}
    \delta\omega
= -\frac{|d_{21}|^2}
              {\pi \hbar \epsilon _0  \mathcal{A}}
               \mathcal{P}
               \int_0^{\infty}
               \! \D\omega\,
               \frac{\omega ^2}{c^2}\,
               \frac{ \mathrm{Im}\,G (z_A, z_A,\omega)}
               {{\tilde \omega}_0 - \omega}.
\end{equation}
Using $\mathrm{Im}G (z_A, z_A,\omega)$ $\!=$ $\![G (z_A, z_A,\omega)
-G (z_A, z_A,\omega)]/(2i)$, we insert the
scattering part of the Green tensor as given by Eq.~(\ref{app2.9})
in Eq.~(\ref{app2.8}).
Then, extending, in rotating-wave approximation,
the integration to $\pm\infty$, we arrive,
after straightforward calculation,
at
Eq.~(\ref{1.43}).


\section{\label{app:2.5}Derivation of Eq. ~(\ref{1.29})}

In the case of a high-$Q$ cavity we are interested in,
we can disregard the contribution to the integral
in Eq.~(\ref{1.40}) of the bulk part of Green function,
i.\,e., we may let
\begin{multline}
  \label{app2.11}
   G (z_A, z_A, \omega )
\mapsto G^S (z_A, z_A, \omega )
\\
    = -
   \sum _k
   \frac{c^2}{\Omega_k |n_1
(\Omega_k)
   |^2 l}
   \frac{1} {\omega-\Omega_k}
   \sin ^2( \omega_k |n_1
(\Omega_k)
   | z_A/c)
\end{multline}
in Eq.~(\ref{1.40}), where we have used Eq.~(\ref{app2.9}).
In this way we obtain
\begin{multline}
  \label{app2.13}
 {\tilde K}(t)
=
        \frac{|d_{21}|^2}
              {\pi \hbar \epsilon _0  \mathcal{A}}
\biggl\{
       \sin ^2( \omega_k |n_1
(\Omega_k)
       | z_A/c)
\bigr.
\\[1ex]
\times
      \int_{\Delta_k}
       \D\omega\,
      \frac{\omega^2}{ 2 i\omega_k |n_1
(\Omega_k)
      |^2 l}
      \frac{1} {\omega-\Omega_k}
       e^{-i (\omega - {\tilde \omega}_0)t}
\\[1ex]
+\sum_{k'\neq k}
 \sin ^2( \omega_{k'} |n_1
(\Omega_k)
 | z_A/c)
\\[1ex]
\Bigl.
\times
    \int_{\Delta_k}
       \D\omega\,
      \frac{\omega ^2}{ 2 i\omega_{k'} |n_1
(\Omega_k)
      |^2 l }
      \frac{1} {\omega-\Omega_{k'}}
       e^{-i (\omega - {\tilde \omega}_0)t}
\biggr\}
.
\end{multline}
Since, within the approximation scheme used, the second
(off-resonant) term may be regarded as being
small comparing to the first one, it can be omitted.
It is then not difficult to prove that, by extending
the $\omega$~integration to $\pm\infty$,
the first (resonant) term can be evaluated to
yield Eq.~(\ref{1.29}).


\section{\label{app:2.7}Derivation of Eq.~(\ref{7.1})}

Using Eqs.~(\ref{app.1}) and  (\ref{app2.17}),
we derive by straightforward calculation
\begin{multline}
  \label{app2.19}
  \frac {\omega ^2} {c^2}
  \int  \D z'\,
  \varepsilon'' (z', \omega)
  G^{(3j')}_{\rm in}(z, z', \omega)
  G^{(1j')*}(z_A, z', \omega)
\\
=
  {\textstyle\frac{1}{2}}
  i  G^{(31)*}(z, z_A, \omega).
\end{multline}
Employing the integral relation
\begin{multline}
  \label{app2.21}
  \frac {\omega ^2} {c^2}
  \int  \D z'\,
  \varepsilon'' (z', \omega)
  G(z_1, z', \omega)G^*(z_2, z', \omega)
\\
=
 \mathrm{Im}\,G
  (z_1, z_2,\omega)
\end{multline}
and using Eqs.~(\ref{app2.15}) and (\ref{app2.19}), we
then
find that
\begin{multline}
  \label{app2.23}
  \frac {\omega ^2} {c^2}
  \int  \D z'\,
  \varepsilon'' (z', \omega)
  G_{\rm out}(z, z', \omega)
  G^{(1j')*}(z_A, z', \omega)
\\
=
  -{\textstyle\frac{1}{2}}
  i  G^{(31)}(z, z_A, \omega).
\end{multline}
Substitution of Eq.~(\ref{1.44}) into Eq.~(\ref{5.11}) and use of
Eq.~(\ref{app2.23}) eventually yield Eq.~(\ref{7.1}).


\section{\label{app:2.9}Validity of
Eq.~(\ref{7.7}) inside the cavity}

Equation~(\ref{1.11}) implies that the
field (in the frequency interval $\Delta_k$)
which propagates inside the cavity from the
left to the right can be written as
\begin{multline}
  \label{app2.31}
    \uh{ E}^{(1)}_{\mathrm{out}}(z,\omega)
      = i \sqrt{\frac {\hbar }{\epsilon _0\pi \mathcal{A}}}\,
      \frac{\omega^2}{c^2}
\\[.5ex]
\times
      \int \D z'\sqrt{\varepsilon''(z',\omega)}\,
      G_{\rm out}(z,z',\omega)
      \hat{f}(z',\omega).
\end{multline}
It is straightforward to prove that the operators
\begin{multline}
  \label{app2.33}
      \hat{ b} ^{(1)}_{\mathrm{out}}(\omega)
    =-2i
    \sqrt{\frac{\varepsilon_0 c \pi \mathcal{A}}
    {\hbar\omega \Gamma_k\gamma_{k\mathrm{rad}}}}
  \,\frac{
   |\omega-\Omega_k|^2
   }
   {\omega-\Omega_k^*}
\\[.5ex]\times
   t_{31}^*(\Omega_k) e^{-i\omega_k l /c}
   \left.\uh{ E} ^{(1)}_{\mathrm{out}}(z,\omega)
       \right|_{z=0_-}
\end{multline}
($n_1$ $\!\equiv$ $\!1$) and
$\hat{ b} ^{(1)\dagger}_{\mathrm{out}}(\omega)$
satisfy the bosonic commutation relation
\begin{equation}
  \label{app2.35}
  \left[\hat{b}^{(1)}_{\mathrm{out}}(\omega),
       \hat{b}^{(1)\dagger}_{\mathrm{out}}(\omega')\right]
       =
\delta (\omega - \omega ') .
\end{equation}
Recalling Eq.~(\ref{5.10}), we derive
\begin{equation}
  \label{app2.37}
  \hat{ b}^{(1)}_{\mathrm{out}}  (\omega)
   |\psi(t)\rangle =
    F^{(1)*}(\omega, t)
   |1\rangle\
   |\lbrace 0\rbrace\rangle,
\end{equation}
where
\begin{equation}
  \label{app2.39}
  F^{(1)}(\omega, t)
  =
  \sqrt{\frac{\Gamma_k} {\gamma_{k\mathrm{rad}}}}
  \,
  F(\omega, t),
\end{equation}
with $F(\omega, t)$ being given by Eq.~(\ref{7.3}).
Similarly to Eq.~(\ref{5.15}), we now
introduce
the unitary transformation
\begin{equation}
  \label{app2.42}
  \hat{b}_{\mathrm{out}\,i}^{(1)}(t)
      = \int_0^{\infty}\!
      \D\omega\,
    F_i^{(1)}(\omega,t)\hat{b}_{\mathrm {out}}^{(1)} (\omega)
\end{equation}
and make the particular choice
\begin{equation}
  \label{app2.40}
  F^{(1)}_1(\omega, t)
  =
  \frac{  F^{(1)}(\omega, t)}
  {\sqrt{\eta^{(1)}(t)}}
\,,
\end{equation}
where
\begin{equation}
  \label{app2.41}
  \eta^{(1)}(t)
  \equiv
  \int_0^{\infty}\!
         \D\omega\, |F^{(1)}(\omega ,t)| ^2
  =
  \frac{\Gamma_k} {\gamma_{k\mathrm{rad}}}\,
    \eta(t).
\end{equation}
Using Eqs.~(\ref{app2.39}) and Eq.~(\ref{app2.41}) and recalling
Eq.~(\ref{5.17}), from Eq.~(\ref{app2.40}) we see that
 \begin{equation}
  \label{app2.45}
  F^{(1)}_1(\omega, t)
 =
   F_1(\omega, t),
\end{equation}
which implies that the
outgoing field
inside the cavity can be described by the same
nonmonochromatic mode
functions $F_i(\omega,t)$
as the excited outgoing field outside the cavity.

Indeed, performing the calculations leading from
Eq.~(\ref{7.41}) to Eq.~(\ref{7.45}) for the outgoing
field inside the cavity
\mbox{($-l$ $\!\leq$ $\!z$ $\!<$ $\!0$)},
\begin{equation}
  \label{app2.46}
    \hat{E}^{(+)}_{\mathrm{out}}(z)
      = \int_0^{\infty}\!
      \D\omega\,
      e^{i\omega z/c}
      \left.
      \uh{E}^{(1)}_{\mathrm {out}} (z, \omega)
\right|_{z=0_-}
,
\end{equation}
instead of the field outside the cavity,
we again arrive at an equation of the form of Eq.~(\ref{7.45}),
with the same
functions
$\phi_i^{(1)}(z,t)$ $\!=$
$\!\phi_i(z,t)$,
but, in general, different associated operators
$\hat{b}_{\mathrm{out}\,i}^{(1)}(t)$
$\!\neq$
$\!\hat{b}_{\mathrm{out}\,i}(t)$. Only in the
limit of vanishing absorption they equal each other.
Performing the calculations leading from Eq.~(\ref{5.7})
to Eq.~(\ref{5.15-1}) for the outgoing field inside the
cavity instead of the outgoing field outside the cavity,
we arrive at the characteristic function
\begin{multline}
  \label{app2.48}
   C_{\mathrm{out}}^{(1)}[\beta_i(t),t]=
    \exp\left[ -{\textstyle\frac{1}{2}}
    \sum_i
      |\beta_i(t)|^2
    \right]
\\[.5ex]
    \times
    \left[
      1-\left|\sum_i \beta_i(t)
     \int_0^{\infty}\!
     \D\omega\,
     F_i^{(1)}(\omega,t)F
^{(1)}
     (\omega, t)\right|^2
    \right]
,
\end{multline}
i.e., $F_i(\omega,t)$ and $F(\omega,t)$ in
Eq.~(\ref{5.15-1}) are simply replaced by
$F_i^{(1)}(\omega,t)$ and $F^{(1)}(\omega,t)$,
respectively, where $F^{(1)}(\omega,t)$ is given
by Eq.~(\ref{app2.39}), and $F_i^{(1)}(\omega,t)$
can be chosen to be $F_i(\omega,t)$.

Needless to say,
that for the operator of the electric field
Eq.~(\ref{1.11})
that refers to
the
points in the region $z$ $\!<$ $\!-l$, one finds
\begin{equation}
  \label{app2.43}
  \uh{ E}^{(0)}_{\mathrm{out}}(z,\omega)
   |\psi(t)\rangle = 0 .
\end{equation}
As expected,
the excited outgoing mode is restricted to the region
$z$ $\!\ge$ $\!-l$.


\section{\label{app:3}Details of Eq.~(\ref{9.31})}

In Eq.~(\ref{9.31}),
$\hat{B}_k(\omega,t)$ is given by
\begin{multline}
  \label{9.35}
  \hat{B}_k(\omega,t) =
      \int _{\Delta_k}
       \D\omega '\,
\Bigl[ G_{k\rm{in}}^* (\omega, \omega ', t)
       \,\hat{b}_{k\mathrm {in}} (\omega ',0)
\Bigr.
\\[1ex]
\Bigl. + \sum_{\lambda} G_{k\lambda}^* (\omega, \omega ', t)
 \,\hat{c} _{k\lambda} (\omega',0)
\Bigr] ,
\end{multline}
where
\begin{multline}
  \label{app3.29}
 G_{k\rm{in}}
        (\omega, \omega ', t) =
    r_{31}^*(\omega)
     e^{i \omega 't}
        \delta (\omega - \omega ')
\\
        -
        T_k^{*2}(\omega)
        \upsilon _k (\omega, \omega ', t)
,
\end{multline}
\begin{equation}
\label{3.31}
    G_{k\rm{cav}}
         (\omega, \omega ', t) =
         -
        T_k ^{*}(\omega) A_{k \mathrm{cav}}^*(\omega)
        \upsilon_k (\omega, \omega ', t),
\end{equation}
\begin{multline}
\label{3.33}
G_{k\pm}
        (\omega, \omega ', t) =
        A^{({\rm o})*}_{k\pm}(\omega) e^{i \omega 't}
        \delta (\omega - \omega ')
\\[1ex]
        -
 T_k ^{*}(\omega) A^*_{k\pm}(\omega)
        \upsilon _k (\omega, \omega ', t)
,
\end{multline}
with
\begin{multline}
\label{app3.35}
        \upsilon (\omega, \omega ', t)  =
        \frac {1}{2\pi}\frac{c}{2n_1 ^*(\omega) l}
        \frac {e^{-i\omega \Delta t}}{
        \omega - \Omega _k ^* }
\\
\times\,
        \left[
        \frac {e^{ i\omega' (t+ \Delta t)}
        - e^{i \Omega _k ^*
        (t+ \Delta t)}}
                {
        \omega' - \Omega _k ^*
        }
\right.
\\
\left.
        -\,
        \frac {e^{i\omega (t+ \Delta t-t_0)}
        - e^{ i\omega' (t+ \Delta t-t_0)}}
                {
        \omega - \omega '
                }
\right].
\end{multline}
In the above,
\begin{align}
  \label{app3.17}
&  A_{k\pm} ^{({\rm o})}(\omega) =
  \frac {t_{23}(\omega)} {
  D_2'(\omega)}
  \frac{1\pm r_{21}(\omega) e^{i \beta _2(\omega) d}}
  {\alpha_{\pm}(\omega)}\,,
\\
 &
A _{ \mathrm{cav}}(\omega ) = -4 i \,\frac{\sqrt{n_1(\omega) }}
   {\alpha_\mathrm{cav}(\omega)} \,,
\\[.5ex]&
  \label{app3.59}
A _{\pm}(\omega ) =
  - \frac {t_{21}(\omega)\sqrt{n_1(\omega)
}} {D_2'(\omega)\alpha_{\pm }(\omega)}
\nonumber\\[.5ex]&\hspace{10ex}\times
  \left[r_{23}(\omega) e^{i \beta _2(\omega) d} \pm 1\right]
  e^{
i \beta _1(\omega) l},
\end{align}
\begin{align}
\label{app.33}
&\alpha_\mathrm{cav} (\omega)
= 2\sqrt{2}|n_1(\omega)|
    \left\lbrace n_1 '(\omega)\sinh[2 \beta _1''(\omega) l]
\right.
\nonumber\\&\hspace{10ex}
\left.
    - n_1''(\omega) \sin[2 \beta _1 '(\omega)l]\right\rbrace^{-\frac{1}{2}} ,
\\[.5ex]
\label{app.35}
&\alpha _{\pm}(\omega)
= |n_2(\omega)|e^{ \beta _2''(\omega) d /2}
    \left\lbrace n_2'(\omega)\sinh[\beta _2''(\omega) d]
\right.
\nonumber\\&\hspace{10ex}
\left.
\pm n_2''(\omega)\sin[\beta _2'(\omega)d]
    \right\rbrace^{-\frac{1}{2}} .
\end{align}

\bibliographystyle{apsrev}
\bibliography{bibl}

\end{document}